\documentclass[twoside,leqno,twocolumn]{article}

\usepackage{ltexpprt}

\usepackage[T1]{fontenc}
\usepackage{lmodern}
\usepackage{amsmath}
\usepackage{amssymb}
\usepackage{graphicx}
\usepackage{booktabs}
\usepackage{tabularx}
\usepackage{array}
\usepackage{hyperref}

\renewcommand{\footnotesize}{\fontsize{8}{10}\selectfont}

\newif\iffullversion
\fullversiontrue % FULL VERSION (arXiv / public repository): appendix included % (build scripts may predefine \buildfull)

\newif\ifcitefullversion
\ifdefined\buildnofull\citefullversionfalse\else\citefullversiontrue\fi % (build scripts may predefine \buildnofull)
\DeclareRobustCommand{\fullver}[1]{\iffullversion Appendix~\ref{#1}\else the full version~\cite{fullversion}\fi}
\DeclareRobustCommand{\Fullver}[1]{\iffullversion Appendix~\ref{#1}\else The full version~\cite{fullversion}\fi}
\DeclareRobustCommand{\withfull}[2]{\iffullversion#1\else\ifcitefullversion#1\else#2\fi\fi}

\newcommand{\Rcal}{\mathcal{R}}
\newcommand{\Scal}{\mathcal{S}}
\newcommand{\Lcal}{\mathcal{L}}

\newcommand{\save}{\operatorname{save}}
\newcommand{\score}{\operatorname{score}}

\newcommand{\code}[1]{\texttt{#1}}

\title{Towards Learning-Guided Discovery of Extremal Geometric\\
Constructions: A PatternBoost Study}
\author{Deepak Ajwani,$^{1}$ Paula Carroll,$^{2}$ Rishikesh Gajjala,$^{4}$
Ananta Manoranjan,$^{4}$\\ Ryan O'Connor,$^{1}$ Rajiv Raman,$^{3}$ Saurabh Ray,$^{5,6}$
and Aimen Taha$^{2}$}
\date{}

\begin{document}
\maketitle
\footnotetext[1]{School of Computer Science, University College Dublin, Ireland}
\footnotetext[2]{Quinn School of Business, University College Dublin, Ireland}
\footnotetext[3]{Indraprastha Institute of Information Technology Delhi, India}
\footnotetext[4]{New York University Abu Dhabi, UAE}
\footnotetext[5]{Computer Science Program, New York University Abu Dhabi, UAE}
\footnotetext[6]{Center for Interdisciplinary Data Science \& AI (CIDSAI), NYU Abu Dhabi, UAE}

\begin{abstract}
Extremal constructions play a central role in the design and analysis of algorithms.  Integrality-gap instances expose the limitations of relaxations, while hard configurations reveal the structural sources of computational difficulty.  Such constructions, however, are still typically discovered by expert researchers through a slow and largely manual process.  This paper investigates how machine learning can accelerate their discovery and thereby support algorithm design and analysis.  We build on PatternBoost, a recent framework that alternates local search with generation from a trained transformer and has shown promise in finding mathematical constructions. We use its Axplorer implementation to investigate three problems: integrality gaps for maximum independent set of rectangles, integrality gaps for stabbing unit squares with axis-parallel lines, and hard instances for recursive guillotine cutting.  By exploring a range of representations of candidate constructions, the local search operations used to modify them, and the surrogate model used to rank candidates before exact verification, we carefully adapt the framework to these combinatorial geometry problems.

While naive representations and search strategies fail to find non-trivial instances, our tailored adaptation discovers numerous non-trivial constructions.  For stabbing unit squares, the method reaches a gap of $20/13\approx 1.54$, which improves on the best known lower bound of $3/2$ in the literature.  Purely automated search converges rapidly and then plateaus, prompting the question of what is required to push the bound higher still.  We find that the diversity of the initial instance set plays an important role.  With modest expert input in the form of a structured staircase construction, the system reaches an improved integrality gap of $14/9\approx 1.555$.  These results demonstrate that machine learning and domain expertise can be combined to discover new extremal constructions that advance algorithmic lower-bound evidence.
\end{abstract}

% Print page numbers (bottom-centred, via ltexpprt's plain page style) and
% number the title/abstract page as 0, so that the body starts at page 1.
% ltexpprt's plain style numbers only odd pages; mirror the centred number
% onto even (left-hand) pages too so every page carries a number.
\pagestyle{plain}
\makeatletter
\let\@evenfoot\@oddfoot
\makeatother

% Place the front matter (title and abstract) on its own first page; per the
% ALENEX submission rules this page does not count toward the 10-page limit.

\section{Introduction}

Extremal constructions are a central driver of progress in algorithm design and analysis.  An integrality-gap instance exposes the limits of a relaxation by separating its value from the true optimum; a hard configuration reveals why a class of algorithms cannot go further.  Such constructions identify the structural obstacles that any successful algorithm must overcome, and they often become the starting point for new theory.  For computational geometry problems, progress has repeatedly depended on a single carefully crafted construction.  Yet these constructions are still usually found by expert researchers through a slow, manual, and problem-specific process.

This paper asks whether machine learning can accelerate the discovery of such extremal objects.  We build on PatternBoost \cite{charton2024patternboost}, a recent framework that alternates local search with generation from a transformer trained on high-scoring candidates, using its Axplorer implementation \cite{axplorer}.  Applying this framework to combinatorial geometry is far from automatic.  Candidate objects have rich geometric structure, and a single combinatorial object has many coordinate encodings, since translating, reflecting, rotating, or slightly perturbing it without changing which objects overlap leaves the objective unchanged. The true objective itself often requires solving a difficult optimization problem; and the learner only succeeds if the representation exposes the relevant combinatorial signal.  We therefore redesign the three components that determine the effectiveness of the search: the representation of candidate objects, the local search operations, and the surrogate model used to rank candidates before exact verification.

We study three problems that capture different forms of extremal construction in combinatorial geometry.  For the maximum independent set of rectangles, we seek instances where a natural relaxation substantially overestimates the size of the largest pairwise disjoint subfamily of rectangles.  For stabbing unit squares with axis-parallel lines, we seek configurations where the fractional stabbing number is significantly smaller than the minimum number of horizontal and vertical lines needed to stab all squares.  For recursive guillotine cutting, we seek configurations that are difficult for algorithms restricted to recursively partitioning the instance by guillotine cuts.

A direct application of the pipeline fails.  Naive representations and generic local search rediscover only trivial constructions or stall at the smallest, easiest gap instances.  After tailoring the pipeline to the geometry and combinatorics of the problems, however, the system discovers non-trivial constructions for all three problems.  For unit-square stabbing, the fully automated system reaches a gap of $20/13\approx 1.54$, exceeding the best known lower bound of $3/2$ for the problem.  For MISR and guillotine cutting, the constructions found ($3/2$ and $1/3$) do not improve on the best known bounds (Section~\ref{sec:problems}); for these two problems the contribution lies in the method and its analysis rather than in new extremal objects.

Our experiments show that this success depends on co-design.  We systematically vary the representation, local search, and surrogate components across the three problems, and independently verify every reported construction.  No single component is sufficient on its own, and the best construction is rare in the design space: for each problem, only one of the $27$ configurations we tried reaches its best value, so obtaining it demands careful exploration of the component choices.

Pure automation, however, has a clear limit.  The search improves rapidly and then plateaus well before its time budget is spent, and simply running it longer does not move it further.  Analyzing the learning dynamics, we find that the plateau is one of reachability: within a fixed search setup the automated loop cannot leave the region it starts in. This leads to a human-in-the-loop strategy: use domain expertise not to hand-design the final construction, but to recast the problem in a structured formulation in which the extremal obstruction is close to native.  For unit-square stabbing this is a staircase construction that lifts a one-dimensional family of intervals to unit squares by products.  With this modest expert input, the unit-square stabbing gap increases further to a certified value of $14/9\approx 1.555$, \emph{improving on the best known lower bound of $3/2$ by a wider margin still}.

These results support a broader view of machine learning as a partner in algorithm engineering.  The role of the model is not to replace mathematical insight, but to amplify it: exploring large construction spaces, identifying promising patterns, and allowing expert guidance to be applied where it is most valuable.

In summary, we contribute:
\begin{enumerate}
\item an adaptation of PatternBoost, through its Axplorer implementation, to extremal construction discovery in combinatorial geometry.  The adaptation includes problem-specific representations, local search operations, and surrogate models.  This turns a pipeline that finds only trivial instances into one that discovers exactly verified constructions, including a $20/13\approx 1.54$ integrality gap for unit-square stabbing that improves on the best known lower bound of $3/2$;
\item a systematic ablation study across three pipeline components and three target problems, with independent verification of every certificate, showing that representation, search, and surrogate ranking must be co-designed: for each problem, only one of the $27$ configurations reaches its best value; and
\item an analysis of the plateau reached by purely automated search, together with a human-in-the-loop strategy based on an expert-structured staircase construction.  This strategy raises the unit-square stabbing gap further to $14/9\approx 1.555$, exceeding the prior best known lower bound and demonstrating productive collaboration between machine learning and domain expertise.
\end{enumerate}

The remainder of the paper is organized as follows.  Section~\ref{sec:related} surveys model-guided construction search, machine learning for combinatorial optimization, and instance-hardness methods.  Section~\ref{sec:problems} defines the three problems and their exact objectives.  Section~\ref{sec:method} describes the PatternBoost/Axplorer framework.  Section~\ref{sec:aggregate} explores the design space of components and presents the constructions found, including the unit-square result that beats the best known bound. Section~\ref{sec:seeding} reports the human-in-the-loop staircase construction.  Section~\ref{sec:discussion} concludes and outlines future directions, and Section~\ref{sec:repro} covers reproducibility.

\section{Related work}
\label{sec:related}

\subsection{Model-guided construction discovery}\ 

A growing body of work uses machine learning to search for explicit mathematical constructions.  Wagner showed that a reinforcement-learning method, the deep cross-entropy method, can produce explicit constructions and counterexamples to conjectures in extremal combinatorics and graph theory \cite{wagner2021}; despite considerable engineering effort, however, we could not get this method to generate non-trivial geometric configurations.  AlphaTensor used deep reinforcement learning to discover matrix-multiplication algorithms that improved on Strassen's construction for the first time in over fifty years \cite{fawzi2022}.  FunSearch pairs a large language model with an automatic evaluator to search over programs, and improved the best known bounds for the cap-set problem and for online bin packing \cite{romeraparedes2024}.  AlphaEvolve extends learning-guided discovery to broader scientific and algorithmic tasks \cite{novikov2025alphaevolve}.  PatternBoost alternates classical local search with a transformer trained on high-scoring candidates \cite{charton2024patternboost}.  We build directly on PatternBoost, through its Axplorer implementation \cite{axplorer}.  In contrast to existing work, our candidates are geometric objects whose exact score is itself an optimization problem, which calls for carefully designed surrogates to guide the search.  A complementary line of work settles extremal questions by complete search, most prominently SAT solving, and emits machine-checkable certificates, as in the resolution of the Boolean Pythagorean triples problem \cite{heule2016}.  We share the insistence on independently verifiable certificates, but replace exhaustive search with learning-guided generation of geometric objects.

\subsection{Machine learning for combinatorial optimization}\ 

This research area has received significant attention in the last decade with a large body of work. We refer the interested reader to the surveys by Bengio et al.\ \cite{bengio2020ml4co}, Cappart et al.\ \cite{cappart2023}, and Wang et al.\ \cite{wang2026ml4co}.  This literature mostly learns to \emph{solve} instances, through learned branching, cutting, or heuristic acceleration, whereas we use learning to \emph{generate} extremal instances.  The generation of extremal instances cannot be easily formulated as a solvable optimization problem: the objective we maximize is itself the output of an exact optimization, an LP and an ILP, so it cannot be predicted directly by a learned model.  While some instance representations can be borrowed from this literature, the model here only proposes candidate instances, each of which is then scored by an exact solver, so every gap we report is an exact, independently computed value rather than a model estimate.

\subsection{Instance hardness and adversarial instance generation}\ 

Searching for configurations that are hard for an exact objective places our task in the instance-hardness and instance-space literature.  In this context, there has been work on instance-space analysis and the evolution of discriminating test instances \cite{smithmiles2015evolving}, principled measures of instance difficulty \cite{smithmiles2012difficulty}, methodologies and open challenges for understanding instance hardness \cite{smithmiles2025hardness}, and the evolution of combinatorial instances that are difficult to solve \cite{vanhemert2006evolving}.  Our work can be read as automated hard-instance discovery, with the distinction that every reported instance carries an exact, independently re-verifiable certificate.

\subsection{Human-guided discovery and diversity-driven search}\ 

Two lines of work inform our human-in-the-loop strategy.  The first is human-guided machine discovery: Davies et al.\ show that machine learning can guide expert mathematical intuition toward new results \cite{davies2021}, and machine discovery and expert refinement can compound, as when human mathematicians rapidly improved on AlphaTensor's machine-found matrix-multiplication schemes \cite{kauersmoosbauer2023}.  Our staircase construction is in the same spirit: the expert does not hand-build the answer but recasts the problem into a structured formulation in which the extremal obstruction is native, and the learned search then tunes it to exceed a known lower bound.  The second is quality-diversity search, which deliberately maintains a diverse population rather than optimizing a single objective: novelty search abandons the objective in favor of behavioral novelty \cite{lehmanstanley2011}, and MAP-Elites illuminates a search space by retaining the best solution in each region of a feature space \cite{mouretclune2015}.  These ideas motivate the diversity controls that let the automated search reach its strongest fully automated construction, and the quality-diversity archive we propose as future work.

\section{Problems and exact objectives}
\label{sec:problems}

For this study, we consider three combinatorial geometry problems -- maximum independent set of rectangles (MISR), unit-square stabbing, and guillotine cutting.  Next, we describe the three problems and the best known upper and lower bounds for them.

\begin{figure*}[t]
\centering
\includegraphics[height=1.8cm]{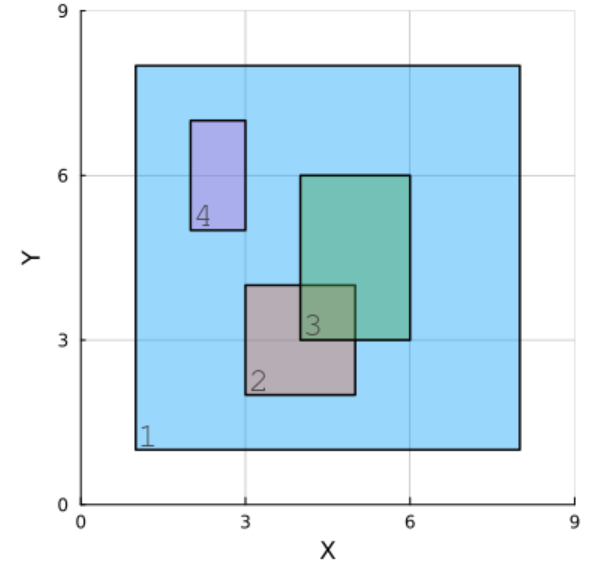}\hfill
\includegraphics[height=1.8cm]{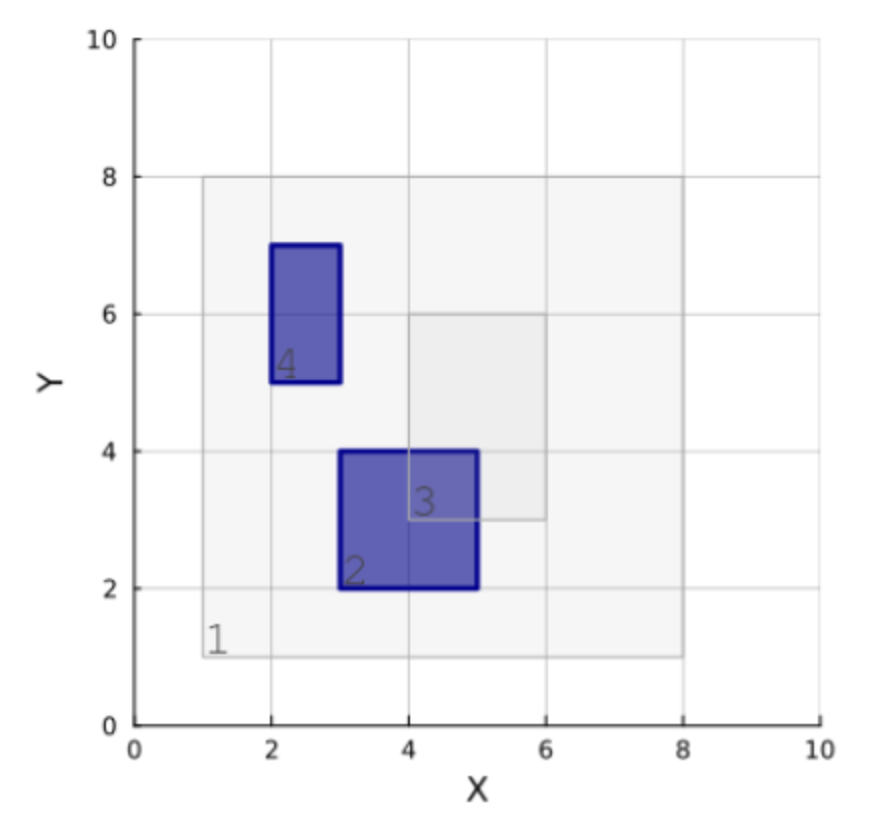}\hfill\hfill
\includegraphics[height=1.8cm]{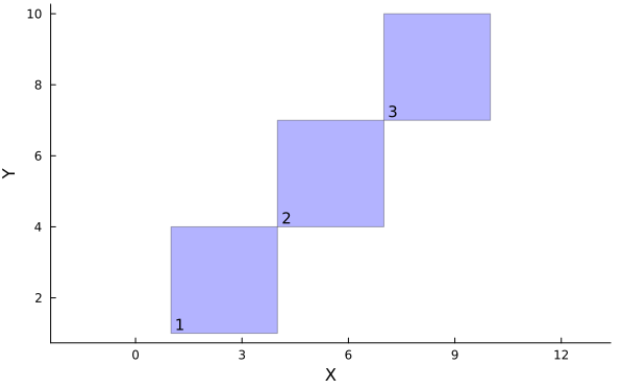}\hfill
\includegraphics[height=1.8cm]{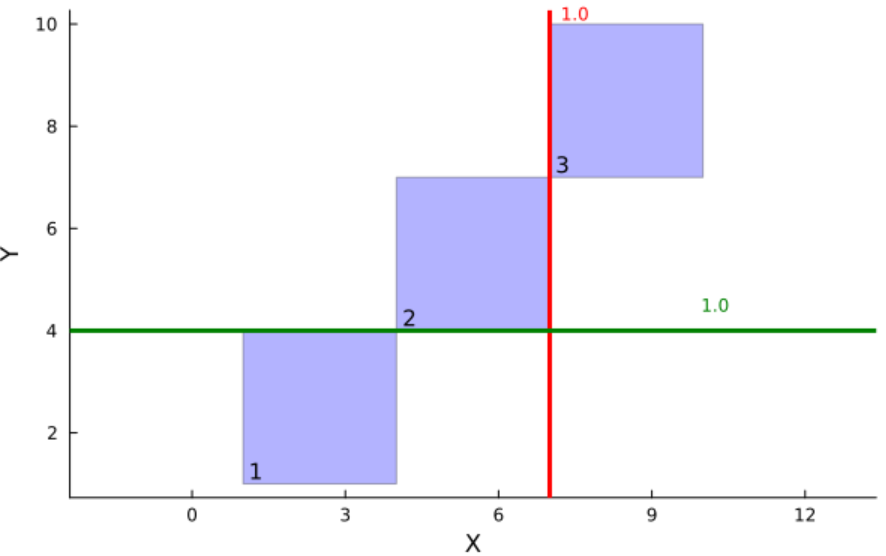}\hfill\hfill
\includegraphics[height=1.8cm]{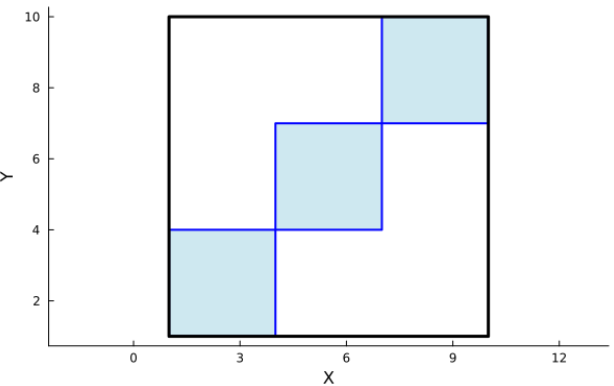}\hfill
\includegraphics[height=1.8cm]{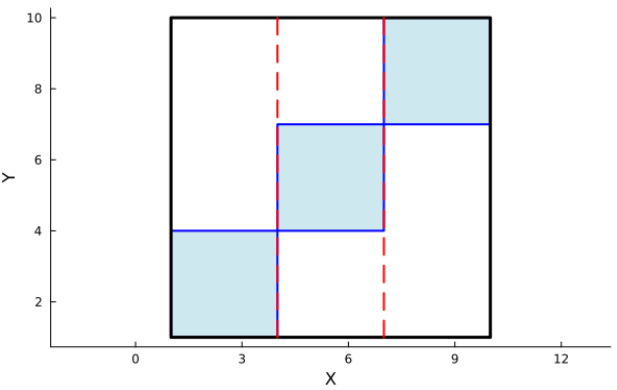}
\caption{Small instances of the three problems, each followed by an optimal solution.  Left pair: A MISR problem instance and a maximum set of pairwise-disjoint rectangles, shaded.  Middle pair: A unit-square stabbing problem instance and an optimal set of horizontal and vertical stabbing lines.  Right pair: A guillotine cut problem instance and the rectangles recovered by an optimal cutting sequence.}
\label{fig:examples}
\end{figure*}

\subsection{MISR integrality gap}\ 

In the MISR problem, we are given a finite family $\Rcal$ of closed axis-parallel rectangles and we seek the maximum number of pairwise-disjoint rectangles.  Let $\alpha(\Rcal)$ denote this integer optimum and let $\alpha^*(\Rcal)$ be the optimum of the standard point (equivalently clique) LP relaxation \cite{chalermsook2009misr}, which assigns $x_R\ge 0$ to each rectangle subject to $\sum_{R\ni p} x_R\le 1$ for every point $p$.  Because axis-parallel rectangles have Helly number two for pairwise intersections, this equals the clique relaxation of the rectangle intersection graph.  The score is
\[
\score_{\mathrm{MISR}}(\Rcal)=\frac{\alpha^*(\Rcal)}{\alpha(\Rcal)} .
\]
Informally, MISR asks for the largest subfamily of pairwise-disjoint rectangles, and a difficult instance is one whose clique-LP optimum $\alpha^*$ substantially exceeds the integer optimum $\alpha$.  Figure~\ref{fig:examples} (left) shows a small instance together with a maximum independent set.

Deciding MISR, and the corresponding covering problem, is NP-complete \cite{fowler1981}.  A long line of work has developed approximation algorithms: Adamaszek and Wiese gave a quasi-polynomial-time approximation scheme (QPTAS) for the weighted problem \cite{adamaszekwiese2013}, and Chuzhoy and Ene obtained an asymptotically faster QPTAS \cite{chuzhoyene2016}, while Mitchell gave the first polynomial-time constant-factor approximation for unweighted MISR \cite{mitchell2021}, later improved to a $3$-approximation by G\'{a}lvez, Khan, Mari, M\"{o}mke, Pittu, and Wiese \cite{galvez2022}.

The quantity we search over is the integrality gap of the clique-LP relaxation, for which lower and upper bounds are known.  Chalermsook and Chuzhoy gave a polynomial-time $O(\log\log n)$-approximation by rounding this relaxation, which also shows that its integrality gap is at most $O(\log\log n)$ \cite{chalermsook2009misr}; a simpler $O(\log\log n)$-approximation was subsequently given by Chalermsook \cite{chalermsook2011}. On the lower side, Chalermsook and Chuzhoy constructed rectangle families whose clique-LP integrality gap is asymptotically at least $3/2$ \cite{chalermsook2009misr}, and Caoduro, Cslovjecsek, Pilipczuk, and W\k{e}grzycki proved essentially tight bounds for axis-parallel segments, exhibiting families whose gap tends to $2$ \cite{caoduro2022independence}.  Since axis-parallel segments are a degenerate special case of axis-parallel rectangles, these families carry over, so the gap for rectangles can likewise be driven arbitrarily close to $2$.  In summary, the best known bounds on the clique-LP integrality gap are an upper bound of $O(\log\log n)$ \cite{chalermsook2009misr} and a lower bound approaching $2$ in the limit \cite{caoduro2022independence}; the value $2$ is a supremum over growing families rather than a gap realized by any single finite instance. Our search targets finite instances of fixed, small size, for which the largest gap realizable is a different and generally smaller quantity than these asymptotic limits; the certificate stores rectangles, the intersection graph, maximal cliques, an optimal integral independent set, LP primal and dual values, and solver status.

\subsection{Unit-square stabbing integrality gap}\ 

In the unit-square stabbing problem, we are given a finite family $\Scal$ of axis-parallel unit squares, and we seek the minimum number of axis-parallel lines that together intersect every square.  A horizontal or vertical line stabs a square if it intersects it.  Let $\Lcal(\Scal)$ be the finite set of critical lines obtained by taking one representative per interval cell induced by the square-side coordinates.  Because the set of squares a line stabs changes only when the line crosses a square edge, all lines whose coordinate lies in the same cell stab exactly the same squares, so this compression is exact: it reduces the infinitely many candidate lines to a finite universe while leaving both the integer and the fractional stabbing optima unchanged.  Let $\tau(\Scal)$ be the minimum number of axis-parallel lines stabbing all squares, and $\tau^*(\Scal)$ the optimum of the set-cover LP over $\Lcal(\Scal)$.  The score is
\[
\score_{\mathrm{stab}}(\Scal)=\frac{\tau(\Scal)}{\tau^*(\Scal)} .
\]
Informally, the goal is to pierce every square with as few horizontal and vertical lines as possible, and a difficult instance is one whose set-cover LP optimum $\tau^*$ is much smaller than the integer optimum $\tau$.  Figure~\ref{fig:examples} (middle) shows a small instance and an optimal integer stabbing.

The rectangle stabbing LP relaxation has integrality gap exactly $2$ for general rectangles \cite{gaur2002,bendavid2012}, and $e/(e-1)\approx 1.582$ for horizontal segments \cite{kovalevaspieksma2006}.  For unit squares, Elbassioni and Ray gave the best known bounds via threshold rounding: the integrality gap is strictly below $2$, at most $1.9\overline{3}$, and at least $3/2$, the lower bound coming from a four-square construction whose LP optimum is $2$ but whose integer optimum is $3$ \cite{elbassioni2024}; they conjecture the true value is $(35-\sqrt{5})/20\approx 1.638$.  The certificate stores square coordinates, the compressed critical-line universe, the incidence matrix, an optimal integer cover, LP primal and dual witnesses, and solver status.

\subsection{Recursive guillotine hardness}\ 

In the guillotine cutting problem, we are given a finite family $\Rcal$ of pairwise interior-disjoint axis-parallel rectangles in a bounding box, and we seek a recursive sequence of guillotine cuts that preserves as many rectangles as possible.  A guillotine cut of a current box is a horizontal or vertical segment spanning the box; rectangles it crosses are destroyed and rectangles strictly on each side recurse.  The recursion continues until every region contains at most one rectangle, and no separate bound on the number of cuts is imposed: since a cut can only destroy the rectangles it crosses, an optimal strategy never uses more than $|\Rcal|-1$ cuts.  Let $\save(\Rcal)$ be the maximum number of rectangles preserved by an optimal recursive strategy.  The score is
\[
\score_{\mathrm{guill}}(\Rcal)=1-\frac{\save(\Rcal)}{|\Rcal|} .
\]
Informally, one recursively slices the bounding box with full horizontal or vertical cuts, and a difficult instance is a packing from which every guillotine sequence must destroy a large fraction of the rectangles.  Figure~\ref{fig:examples} (right) shows a small instance and the outcome of an optimal cutting sequence.

Guillotine separability and cutting sequences were studied by Pach and Tardos \cite{pachtardos2000}, Abed et al.~\cite{abed2015}, and Khan and Pittu \cite{khanpittu2020}.  The central quantity is the fraction of rectangles a guillotine sequence can separate.  For general disjoint rectangles, Pach and Tardos showed that an $\Omega(1/\log n)$ fraction can always be separated and conjectured that a constant fraction always suffices \cite{pachtardos2000}; this remains open.  For squares the separable fraction is at least a constant: Abed et al.\ showed that some constant fraction can always be separated~\cite{abed2015}, and the best explicit guarantee was improved to $1/40$~\cite{khanpittu2020} and then to $9/256\approx 3.51\%$~\cite{chalermsook2025squares}.  In the other direction, there is an arrangement of squares from which no guillotine sequence can separate more than half.  The optimal constant for squares therefore lies between $9/256$ and $1/2$.  Such separability bounds translate directly into dynamic-programming approximations for the corresponding independent-set problems.  The exact verifier is a dynamic program over coordinate-induced rectangle-subset states; the certificate stores saved and destroyed counts, the optimal recursive cut tree, first-cut and threshold diagnostics, the DP state count, and solver status.

Each problem is scored by an exact objective computed by a verifier that is independent of the search, so that the exact score, and not a surrogate, is the value reported everywhere.\withfull{  \Fullver{app:examples} gives small, easily verifiable instances that certify a non-trivial objective for each problem: an integrality gap above $1$ for MISR and unit-square stabbing, and a destroyed fraction above $0$ for guillotine cutting.}{}

\section{PatternBoost/Axplorer framework}
\label{sec:method}

\begin{figure}[t]
\centering
\includegraphics[width=0.62\columnwidth]{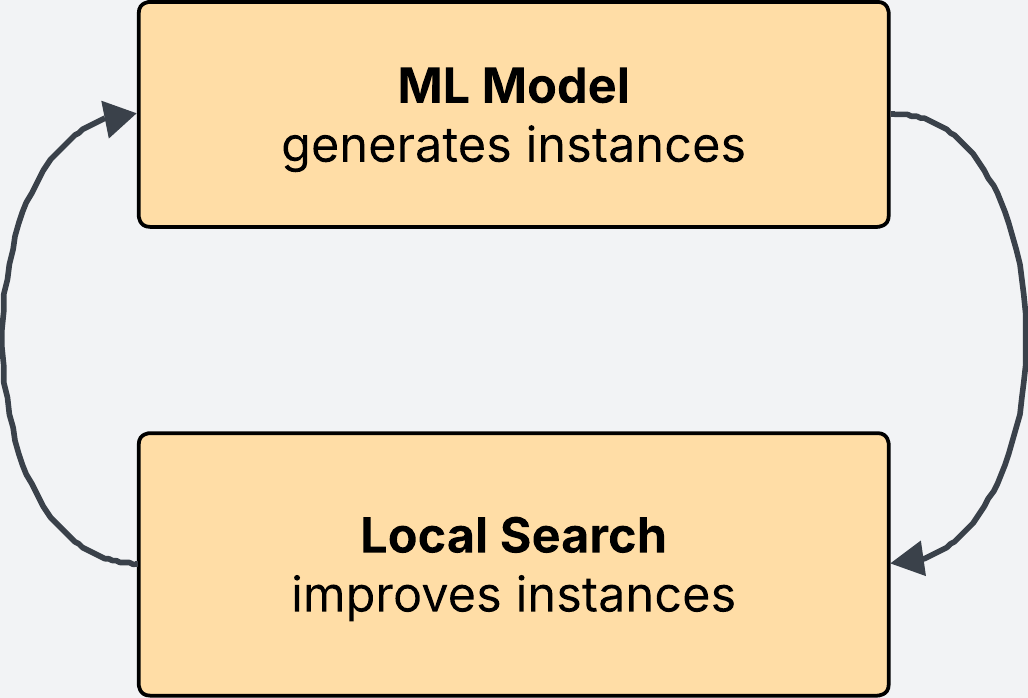}
\caption{The PatternBoost loop: local search and transformer generation alternate over a database of high-scoring constructions.  Adapted from \cite{charton2024patternboost}.}
\label{fig:pbloop}
\end{figure}

\paragraph{The PatternBoost/Axplorer loop.}\ 
PatternBoost \cite{charton2024patternboost} was recently proposed as a framework for discovering mathematical constructions with the help of a trained model.  It alternates two phases over a growing database of scored constructions.  In the \emph{local-search} phase, existing constructions are perturbed by small targeted moves and improved ones are kept; in the \emph{global-learning} phase, a transformer trained on the highest-scoring constructions in the database generates new candidates, which are handed back to local search.  The transformer supplies globally informed starting points that local search alone would rarely reach, while local search supplies the correction that turns rough generated candidates into valid constructions; the combination outperforms either phase alone \cite{charton2024patternboost}.  In principle, only the problem-specific scoring function needs to change between problems.  However, we discovered in this work that the local search operation and the representation also need to be carefully engineered for each problem.  Figure~\ref{fig:pbloop} shows the loop\withfull{; pseudocode is given in \fullver{app:pseudocode}}{, and its implementation is in the artifact repository (Section~\ref{sec:repro})}.  It is a population-based search: each generation scores the population with the surrogate, exactly audits the elites at a fixed interval, periodically retrains a fresh model on the archive, and refills the population by mutating elites with the configuration's local-search operator and adding a few random newcomers.

\emph{The transformer.}  A construction is serialized as a canonical JSON string and tokenized character by character, with start and end markers.  The model is a decoder-only transformer that predicts the next token autoregressively and is trained by next-token cross-entropy on the elite database, so it learns the token patterns shared by high-scoring constructions.  We deliberately keep it small and train it from scratch at every training event: two layers, four attention heads, embedding width $96$, feed-forward width $384$, GELU activation, no dropout, a causal attention mask, context length $128$, and generated sequences of up to $512$ tokens.  It is optimized with AdamW at learning rate $3\times10^{-4}$, batch size $32$, and three epochs per training call.  The vocabulary is rebuilt from the current elite archive with explicit \code{<BOS>}, \code{<EOS>}, and \code{<UNK>} symbols, so there is no fixed global vocabulary size.  Character-level tokenization keeps the vocabulary small at the cost of allowing malformed samples, which are discarded or repaired; when a training event yields too few parseable samples, an order-$8$ character $n$-gram model fitted to the same archive supplements the transformer output.

\emph{The surrogate.}  Each configuration carries a surrogate: a problem-specific function that ranks candidates within the search and selects the elite that will be exactly audited.  Surrogates range from cheap structural statistics that never solve the objective, through approximations of it, to functions that invoke the exact scorer itself with additional shaping terms.  A surrogate is therefore not necessarily an approximation: what separates it from the exact oracle is its role, since the oracle sits outside the search and alone determines reported scores.

\emph{Local search and verification.}  A local move makes a small, validity-preserving change to a construction, in the simplest case perturbing one coordinate by $\pm1$.  The operations we use range from these simple coordinate perturbations to structure-aware moves and small, exactly scored hill climbs; the concrete operators considered for each problem are described in Section~\ref{sec:aggregate}.  Improved candidates are accepted greedily.

\paragraph{Challenges in adapting to combinatorial geometry.}\ 
Adapting PatternBoost to combinatorial geometry is not automatic, because the geometric setting departs from the graph and sequence problems the framework was designed for in three ways.  First, a construction is a set of geometric objects with large coordinate ranges and many symmetries -- translation, reflection, and axis exchange -- so a naive coordinate encoding represents each combinatorial structure in a great many equivalent ways and forces the transformer to spend its capacity relearning them.  Second, the objective is not a closed-form count but the outcome of an exact optimization -- an LP, an ILP and their ratio, or a dynamic program -- which is expensive to evaluate and, being a ratio, changes only in coarse jumps, so most local moves leave it unchanged.  Third, randomly generated or model-sampled coordinates rarely satisfy the validity constraints of a non-trivial instance, such as disjointness, triangle-freeness, or congruence, so proposals must be repaired, and repair can silently undo a coordinated change.  These difficulties are why the representation, the local-search operation, and the surrogate must be co-designed for each problem, which we take up in Section~\ref{sec:aggregate}.

\paragraph{Exact verification and certificates.}\ 
Trust rests on verification that is independent of the search.  Before deduplication and scoring, candidates are canonicalized: coordinates are translated to a canonical origin, integer grids are scaled by their gcd when possible, objects are sorted by normalized coordinates or a canonical graph label, and ties are broken deterministically.  Surrogates are allowed to be biased and problem-specific, but every reported value is recomputed by the exact oracle of Section~\ref{sec:problems} and saved as a standalone certificate.  These oracles use only open-source solvers, with no commercial optimization license; the per-problem solvers and numerical settings are given in Section~\ref{sec:setup}.  To keep results non-trivial, accepted constructions required at least eight geometric objects, unit-square certificates required integer stabbing number at least four, and guillotine certificates required at least two destroyed rectangles.  We disclose the strongest prior: one of the three guillotine representations builds its instances around a fixed seven-rectangle spiral obstruction topology, so the nine configurations using it search an obstruction family rather than neutral coordinates.  It inserts neither a score nor a certificate, and our best guillotine construction does not come from it: the destroyed fraction of $1/3$ is reached by the neutral \code{rect\_direct\_disjoint} representation, while the spiral-seeded configurations reach at most $0.27$. A strict post-processing audit re-verified every construction, checking clean completion of every run, agreement among all recorded scores, certificate-hash consistency, and independent exact re-verification of every certificate.  All runs terminated cleanly, and all $81$ constructions passing the audit.

\section{Exploring the design space of components for geometric construction discovery}
\label{sec:aggregate}

\subsection{Components}
\label{sec:components}

\paragraph{The components and how we explore them.}\ 
The effectiveness of the search is governed by three components: the \emph{representation}, which specifies how a construction is encoded and repaired; the \emph{local-search operation}, which specifies how an encoded construction is changed; and the \emph{surrogate}, which scores how promising each candidate is and
ranks candidates for exact verification. For each problem we consider three choices of each component and explore every one of the resulting $27$ combinations.  Guided by the principle of shaping each component to the certificate being sought, the choices for each problem range from direct coordinate encodings to structured encodings that build the target property into generation, from simple coordinate perturbations to exactly scored, structure-aware hill climbs, and from cheap graph or incidence statistics to surrogates that invoke the exact scorer during ranking.  Table~\ref{tab:components} lists all $27$ choices.

\begin{table*}[t]
\caption{The $27$ component choices, three per component and problem.  The components of each problem's best configuration are in bold\withfull{; \fullver{app:components-detail} describes every component in detail}{}.}
\label{tab:components}
\centering
\footnotesize
\newcommand{\seqcode}[1]{\begingroup\def\_{\char`\_\discretionary{}{}{}}\texttt{#1}\endgroup}
\renewcommand{\code}[1]{\seqcode{#1}}
\begin{tabularx}{\textwidth}{@{}p{0.075\textwidth}>{\raggedright\arraybackslash}X>{\raggedright\arraybackslash}X>{\raggedright\arraybackslash}X@{}}
\toprule
Problem & Representation & Local-search operation & Surrogate \\
\midrule
MISR & \textbf{\code{triangle\_free\_rect}}: direct integer rectangles; generation and repair keep the intersection graph triangle-free. & \textbf{\code{sequence\_pair\_pivot}}: representation-safe add, replace, nudge, resize, and delete moves, with add/replace upweighted. & \textbf{\code{exact\_lp\_gap\_pressure}}: the exact gap $\alpha^*/\alpha$, with a triangle-free lower-bound tie-break. \\
 & \code{quadratic\_program\_rectangles}: bounded integer coefficient program decoded to rectangles, then repaired to be triangle-free. & \code{program\_coeff\_pivot}: perturbs coefficient blocks, decoder choice, or size of a program payload; else safe direct moves. & \code{triangle\_free\_exact\_gap\_pressure}: rejects candidates with a triangle, then ranks by exact gap. \\
 & \code{fixed\_symmetry\_rectangles}: direct integer rectangles at a fixed count, canonicalized under translation, reflection, and axis exchange. & \code{symmetry\_crossover\_hillclimb}: $25$ exactly scored one-step shifts, relocations, coordinate exchanges, and symmetry moves; keeps the best non-worsening one. & \code{graph\_conflict\_proxy}: cheap graph statistic based on intersection-edge density, independence number, clique size, and a triangle-free bonus.\\
\addlinespace
Unit-square & \textbf{\code{sqstab\_exact\_grid}}: distinct integer anchor endpoints plus one common side length, normalized to a canonical origin. & \code{coord\_mutation}: moves one square by one grid step, duplicates, or deletes it, then repairs. & \textbf{\code{exact\_stab\_gap\_pressure}}: the exact gap $\tau/\tau^*$, with tie-breaks for a large integer cover, low LP value, and non-trivial LP support. \\
 & \code{line\_square\_incidence}: a balanced abstract row/column incidence layout mapped to square coordinates. & \code{sqstab\_local\_hillclimb}: stabbing-aware line-tightening, add/replace, delete, restart, and shift candidates; keeps the best. & \code{greedy\_partial\_lp\_bitset}: greedy cover size over a max-line-frequency fractional bound, from incidence bitsets. \\
 & \code{fixed\_symmetry\_grid}: distinct anchor endpoints at a fixed side and count, canonicalized under planar symmetries. & \textbf{\code{symmetry\_crossover\_hillclimb}}: same symmetry\_crossover\_hillclimb described above. & \code{incidence\_statistics}: uses greedy cover size and balance of line frequencies; a cheap incidence-only proxy. \\
\addlinespace
Guillotine & \textbf{\code{rect\_direct\_disjoint}}: direct disjoint rectangles; repair removes overlaps and enforces size bounds. & \code{packing\_resize}: moves, resizes, adds, or replaces rectangles, then applies nonseparability-guided step. & \textbf{\code{depth\_limited\_dp}}: a depth-four guillotine dynamic-programming estimate of the destroyed fraction. \\
 & \code{recursive\_obstruction\_grammar}: random seven-rectangle spiral obstruction cores plus filler rectangles; a random generator
family. & \textbf{\code{witness\_breaking}}: finds a guillotine-separable majority subset and adds a bridge or blocker that breaks it. & \code{first\_cut\_obstruction}: measures number of rectangles crossed by the best first cut, with balance and span bonuses. \\
 & \code{fixed\_symmetry\_packing}: a fixed-count random disjoint rectangle packing, canonicalized under symmetries. & \code{symmetry\_crossover\_hillclimb}: same symmetry\_crossover\_hillclimb described above. & \code{k\_subset\_nonseparability}: fraction of majority subsets resisting guillotine separation, with shallow-DP and beam terms. \\
\bottomrule
\end{tabularx}
\end{table*}

\paragraph{Components of the best configurations.}\ 
We describe in more detail the components that make up the best configuration for each problem.

For MISR, \code{triangle\_free\_rect} stores a list of integer rectangle coordinates under the constraint that the intersection graph stay triangle-free, that is, no three rectangles share a common point.  Rectangles are added one at a time from a pool that mixes rectangles near existing ones, thin horizontal and vertical bars, and unrestricted random rectangles; a candidate is admissible only if its neighbors in the current intersection graph are pairwise non-intersecting, which is exactly what prevents a triangle.  Repair deletes a member of any triangle that appears and refills to the minimum size.  On this representation, \code{sequence\_pair\_pivot} (the local search) samples add, replace, nudge, resize, and delete moves with add and replace upweighted, drawing new rectangles from a large triangle-free pool.  The surrogate \code{exact\_lp\_gap\_pressure} calls the full MISR exact scorer for every candidate it ranks and orders the population primarily by the exact clique-LP-to-integer gap $\alpha^*/\alpha$, breaking ties in favor of a larger triangle-free lower bound.  It is the most expensive MISR signal, since each ranking call solves the optimization exactly, same as the verifier.

For unit-square stabbing, \code{sqstab\_exact\_grid} stores distinct integer lower-left anchor endpoints and one common integer side length. Canonicalization removes translation and duplicates, so the exact critical-line universe is easy to regenerate and audit.  \code{symmetry\_crossover\_hillclimb} is an operator shared by all three problems that makes $25$ exactly scored proposals per call, drawn as about $55\%$ one-step shifts, $25\%$ random relocations, $10\%$ pairwise exchanges (swapping the vertical intervals of two objects), and $10\%$ planar symmetries.  Every proposal is repaired and exactly scored; acceptance is gated first by a non-triviality requirement (here, a candidate needing at least four lines is preferred) and then by the exact objective, and only non-worsening proposals are kept.  \code{exact\_stab\_gap\_pressure} calls the exact set-cover and fractional-LP scorer for each candidate and ranks primarily by $\tau/\tau^*$, with small tie-breaks that prefer a larger integer cover, a smaller fractional optimum, and a more spread-out LP support.

For guillotine cut, \code{rect\_direct\_disjoint} initializes by turning random grid cells into disjoint unit rectangles, a deliberately neutral start that usually yields an easily separable packing; repair removes interior overlaps, refills to the minimum size, and canonicalizes, leaving the local search to grow the simple cells into long bars and interlocking packings.  \code{witness\_breaking} first finds an easy witness, a guillotine-separable subset of majority size (enumerated for small $n$, sampled otherwise), and then tries to destroy that separation by extending a rectangle across a projection gap or inserting a thin bridging rectangle, alongside bar replacement, cut blockers, and motif copies.  \code{depth\_limited\_dp} runs the recursive-cut dynamic program to depth four, optimistically treating every rectangle still present at the depth limit as saved, and ranks by the implied destroyed fraction; it sees past the first cut while staying cheaper than the full dynamic program, but misses obstructions that surface deeper than four levels. 

\subsection{Experimental setup}
\label{sec:setup}

\paragraph{Experimental design.}\ 
For each problem the three components are crossed fully, giving $27$ configurations, and the three problems together give $81$.  Each configuration repeatedly generates or repairs a population in its representation, scores novel candidates with its surrogate, exactly audits the elite at a fixed interval and saves a certificate only when the exact score strictly improves, mutates the elite with its local-search operation, and periodically trains a freshly initialized transformer on the strongest archive entries and samples new candidates from it, with a character $n$-gram sampler filling any shortfall of parseable samples.  Every configuration is run once, from freshly generated geometry, with no repeated-seed axis and no warm-started certificate, under an identical 24-hour wall-clock budget.  To attribute an effect to a single component we use its \emph{marginal mean}, the average exact score over the $9$ configurations that share that component choice. Because there is one run per configuration, these marginals are descriptive summaries rather than estimates with error bars.

\paragraph{Hardware and run cohorts.}\ 
Experiments ran on compute nodes built from AMD EPYC~7742 processors at $2.25$~GHz with $128$ cores and $480$~GB of RAM each (about $3.75$~GB per core). Each configuration was dispatched as one array task requesting four cores and $16$~GB of RAM.  All configurations shared exact elite scoring every five generations, $16$ requested model samples per successful training event, per-generation checkpointing, an application budget of $86{,}400$~seconds, and a safety cap of $1{,}000{,}000$ generations; the runs were thus both time-bounded and iteration-capped, but every reported configuration stopped on the time budget and none reached the iteration cap.  The set of $81$ configurations combines three runtime cohorts rather than one common population and elite setting: $36$ rows used object count $n=8$ on a grid of $8$ with population $32$, elite $6$, and training every seven generations; $42$ rows used $n=12$ on grid $8$ for MISR and guillotine and $n=20$ on grid $16$ for unit-square stabbing, with population $16$, elite $4$, and training every ten generations; and the final three unit-square rows used $n=20$ on grid $16$ with population $32$, elite $12$, an initial pool of $128$ candidates, and four fresh immigrants per generation.

\paragraph{Initialization.}\ 
Every run starts from fresh pseudorandom geometry rather than an injected certificate, and the random seeds are retained only for provenance and replay.  Initialization is problem- and representation-specific.  MISR uses integral rectangle coordinates: at grid $8$, repaired lower endpoints lie in $0$--$9$ and upper endpoints in $1$--$10$, and the quadratic-program representation draws its initial coefficients from $[-4,4]$.  Unit-square instances place congruent squares whose anchors lie in $0$--$8$ at grid $8$ and $0$--$16$ at grid $16$; the exact-grid representation samples a common integer side from $1$ to $4$, while the fixed-symmetry representation fixes the side at $2$.  Guillotine populations in the direct representation choose disjoint unit grid squares from an $8\times8$ grid, the random fixed packing uses a dynamic domain of side at least $\max(\text{grid}+2,\,2n)$, and the recursive-obstruction representation randomizes core spacings and orientation and adds filler rectangles.  Initial object counts are $8$ in the retained cohort, $12$ for replacement MISR and guillotine, and $20$ for replacement unit-square stabbing, and mutation and repair keep the count within $8$--$24$, $12$--$36$, and $20$--$60$ respectively, except for representations that explicitly preserve cardinality.

\paragraph{Progress axes and metrics.}\ 
We retain three progress axes because they answer different questions: search generation (the most direct algorithmic-progress variable), elapsed time (the fairest resource comparison, since every configuration received the same wall-clock budget), and cumulative transformer-training epoch (an accounting axis over repeated short fits, not a single continuous learning curve).  Per-configuration metrics are the best exact score, time-to-best, epoch-to-best, and the post-best plateau fraction.

\paragraph{Exact solvers.}\ 
The exact oracles that gate certificate acceptance use only open-source solvers, with no commercial optimization license.  MISR combines Bron--Kerbosch maximal-clique enumeration, SciPy's HiGHS linear-programming backend for the fractional clique LP, and a custom bitset branch-and-bound maximum-independent-set solver; unit-square stabbing enumerates the finite critical-line universe, uses the same HiGHS backend for the fractional cover, and a custom branch-and-bound set-cover solver for the integer optimum; and guillotine hardness uses a custom recursive dynamic program over rectangle-subset bitmasks and all relevant cuts.  Each linear program runs on a single HiGHS thread with no separate solver time limit, verification uses a numerical tolerance of $10^{-8}$, and the guillotine dynamic program uses exact discrete comparisons.

\subsection{Results}
\label{sec:results}

\paragraph{Good solutions are rare.}\ 
Exploring this space succeeds, but only sparsely.  Table~\ref{tab:best} reports, for each problem, the best configuration and its exact score together with the next-best and median scores over the $27$ configurations. The complete per-configuration results, including the gain over the initial construction, the training epoch and elapsed time at which the best appeared, and the post-best plateau fraction, are given in \withfull{\fullver{app:all81}}{the artifact repository (Section~\ref{sec:repro})}.  The weakest configurations, which pair a generic coordinate representation with a cheap heuristic surrogate, never move beyond trivial or half-integral constructions, and for every problem the maximum is reached by a single configuration out of the $27$.  Good solutions are therefore rare: they occupy isolated points of the design space rather than a broad basin, and the median configuration lies well below the maximum (Table~\ref{tab:best}).

\newcommand{\tinyscriptsize}{\fontsize{5.7pt}{7.8pt}\selectfont}

\begin{table}[t]
\caption{Best configuration for each problem among its $27$. Best is the exact score ($\alpha^*/\alpha$, $\tau/\tau^*$, and $1-\save/n$, respectively); Next is the highest score among the other $26$ configurations; Median is the median over all $27$; $n$ is the object count and h the elapsed hours at the best construction. Complete per-configuration results are given in \withfull{\fullver{app:all81}}{the artifact repository (Section~\ref{sec:repro})}.}
\label{tab:best}
\centering
\small
\newcommand{\seqcode}[1]{\begingroup\def\_{\char`\_\discretionary{}{}{}}\texttt{#1}\endgroup}
\renewcommand{\code}[1]{\seqcode{#1}}
\begin{tabular}{@{}lccccr@{}}
\toprule
Problem & Best & Next & Median & $n$ & h \\
\midrule
MISR & $3/2$ & $1.400$ & $1.300$ & $12$ & $3.12$ \\
\multicolumn{6}{@{}>{\centering\arraybackslash}p{0.95\columnwidth}@{}}{\tinyscriptsize\code{triangle\_free\_rect} / \code{sequence\_pair\_pivot} / \code{exact\_lp\_gap\_pressure}} \\
\addlinespace
Unit-square & $20/13$ & $1.500$ & $1.414$ & $20$ & $0.37$ \\
\multicolumn{6}{@{}>{\centering\arraybackslash}p{0.95\columnwidth}@{}}{\tinyscriptsize\code{sqstab\_exact\_grid} / \code{symmetry\_crossover\_hillclimb} / \code{exact\_stab\_gap\_pressure}} \\
\addlinespace
Guillotine & $1/3$ & $0.273$ & $0.250$ & $9$ & $5.81$ \\
\multicolumn{6}{@{}>{\centering\arraybackslash}p{0.95\columnwidth}@{}}{\tinyscriptsize\code{rect\_direct\_disjoint} / \code{witness\_breaking} / \code{depth\_limited\_dp}} \\
\bottomrule
\end{tabular}
\end{table}

\begin{figure}[t]
\centering
\includegraphics[width=0.55\columnwidth]{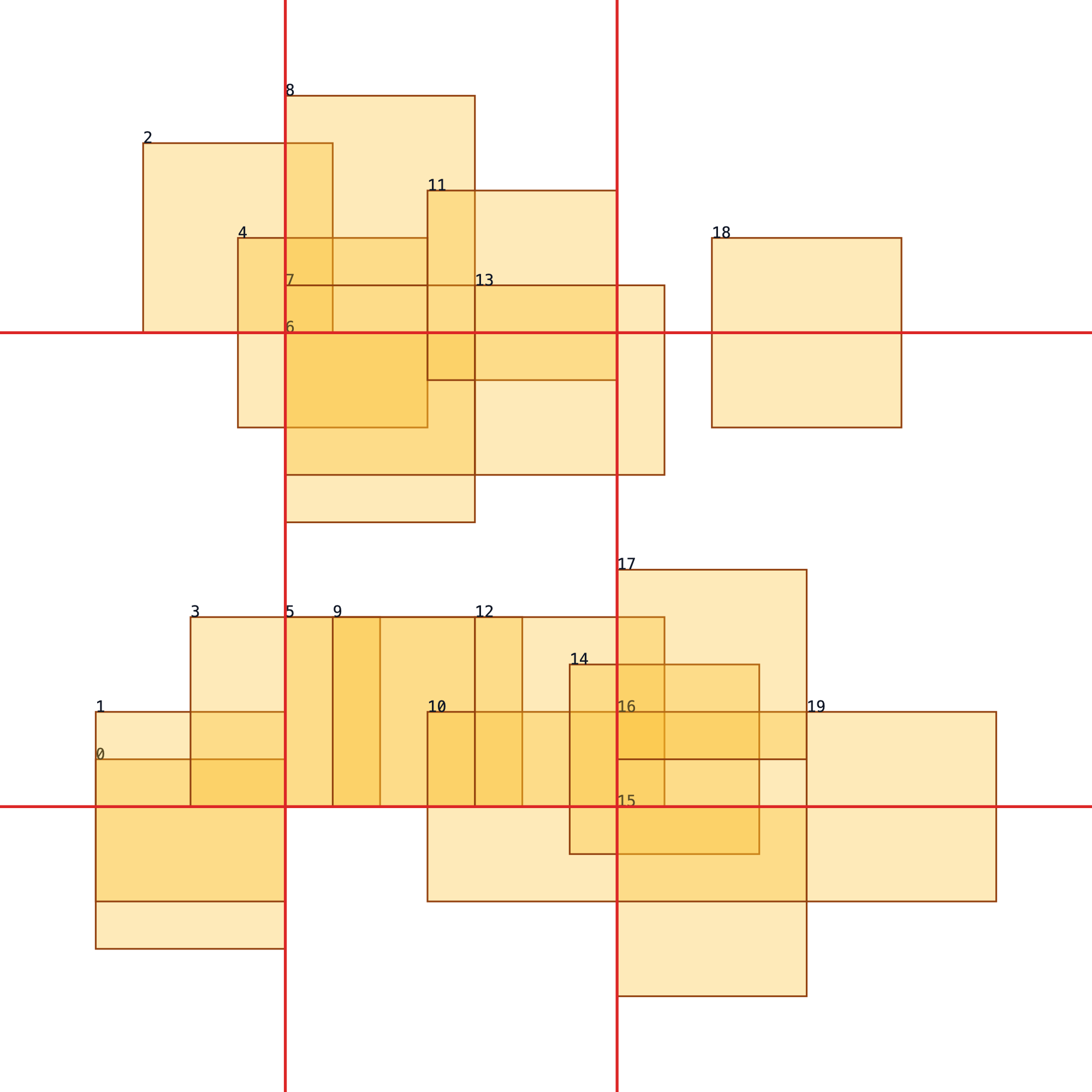}
\caption{Best audited unit-square stabbing construction found by the automated search.  The $20$ congruent squares (labeled $0$--$19$) admit an optimal integer stabbing cover of $\tau=4$ lines (red) while the set-cover LP optimum is $\tau^*=13/5$, giving the integrality gap $20/13\approx 1.54$.}
\label{fig:sqstb-record}
\end{figure}

The best constructions are small, exactly verified witnesses.  For MISR the search attains an integrality gap of $3/2$: a family of twelve rectangles with clique-LP optimum $\alpha^*=6$ and integer optimum $\alpha=4$.  For guillotine hardness it attains a destroyed fraction of $1/3$: a packing of $9$ rectangles from which an optimal recursive strategy preserves six and destroys three, found from scratch with no warm start.  The strongest single outcome is for unit-square stabbing, where the search attains an integrality gap of $20/13\approx 1.54$: a family of twenty congruent squares with set-cover LP optimum $\tau^*=13/5$ against an integer optimum $\tau=4$ (Figure~\ref{fig:sqstb-record}).  This is a major achievement, since it strictly exceeds the best lower bound of $3/2$ previously known in the literature (Section~\ref{sec:problems}) and is found automatically with an independently verified certificate.  It is also a rare single-configuration outcome: of the $27$ unit-square configurations, $9$ reach exactly $3/2$ and only this one exceeds it.

\begin{figure*}[t]
\centering
\includegraphics[width=\textwidth]{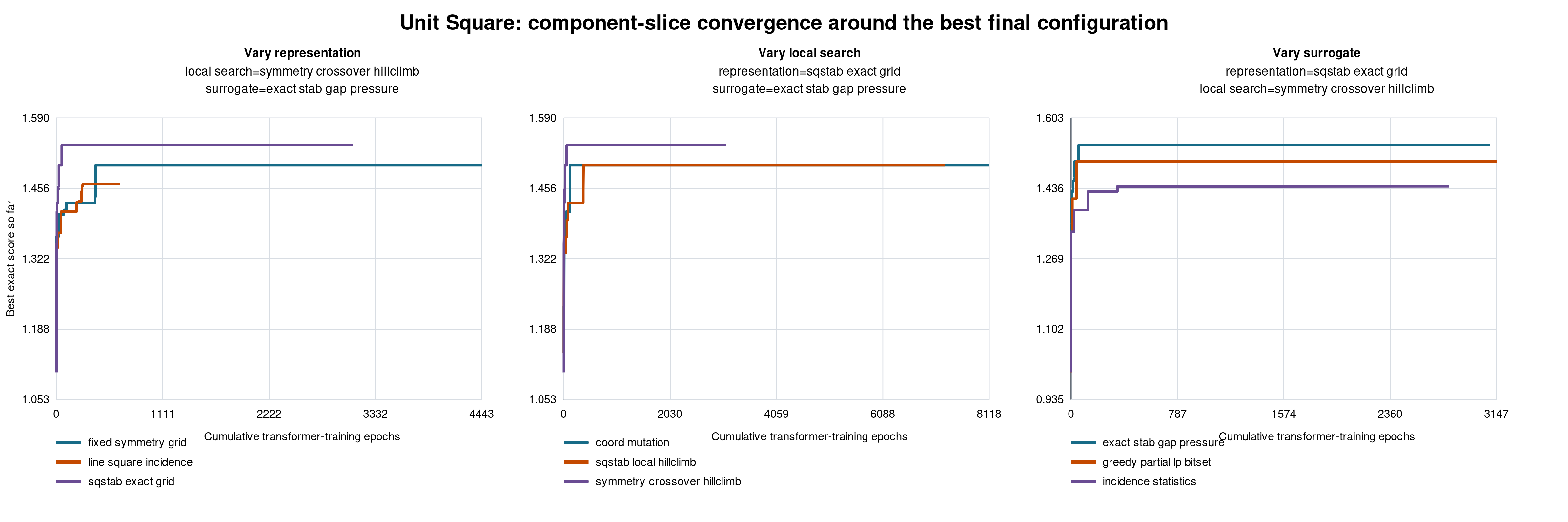}
\caption{Component-slice convergence for unit-square stabbing.  Each panel fixes two of the three components at their best-performing choice and varies the third, plotting the best exact integrality gap $\tau/\tau^*$ found so far against cumulative transformer-training epochs.  From left to right the panels vary the representation, the local-search operation, and the surrogate; the surrogate panel shows the strongest single-component separation in the study.}
\label{fig:slice-unitsquare}
\end{figure*}

\paragraph{Which combination unlocks each problem.}\ 
The winning configuration is problem-specific.  The best combinations (Table~\ref{tab:best}) keep the rectangle intersection graph sparse and score by the true LP/integer gap for MISR, operate on the line--square incidence guided by the exact stabbing gap for unit-square stabbing, and attack the recursive cut that would otherwise separate the packing for guillotine hardness.  Figure~\ref{fig:slice-unitsquare} illustrates the effect component by component for unit-square stabbing: holding two components at their best choice and varying the third shows the surrogate as the sharpest lever, with the exact stabbing-gap surrogate climbing well above the cheap proxies.\withfull{  The analogous slices for MISR and guillotine hardness are given in \fullver{app:slices}.}{}

\paragraph{Careful exploration and co-design.}\ 
Taken together, the exploration shows that no component is decisive on its own and that the component carrying a problem's structural signal differs across problems: a representation that preserves sparse conflicts for MISR, a surrogate that measures the true gap for stabbing, and a mutation that attacks the recursive escape route for guillotine hardness.  Because good solutions occupy isolated points of the design space, they are reached only when the representation, the local-search operation, and the surrogate are chosen together for the problem at hand; a mismatch in any one of them caps the achievable score, however well the other two are tuned.  This careful, joint exploration of the components, rather than the learned model on its own, is what turns the generic loop into a working discovery method.  The records nonetheless stop improving well before the budget is exhausted, for reasons we examine next.

\paragraph{The plateau of automated search.}\
The best scores reported above are reached quickly and then stop improving.  We ask whether this plateau reflects a shortage of computation or a structural limit, and three measurements point to the latter.  First, the ranking of configurations stabilizes early: by one quarter of the training budget, the Spearman rank correlation between the current ordering of the $27$ configurations and their final ordering is already $0.908$ for MISR, $0.899$ for unit-square, and $0.944$ for guillotine, where a value of $1$ would mean the ordering never changes again; the remaining runtime thus mostly extends plateaus rather than rearranging which components win. Second, the learned model rarely produces a record: counting only sampled candidates that pass exact verification and strictly improve the best score, the transformer accounts for $0.0\%$, $0.5\%$, and $1.5\%$ of all improvements ($0$, $1$, and $1$ records), against $78$, $220$, and $63$ from the combined initialization and local-search path.  Third, the plateaus are long: in the median configuration, $97.5\%$, $64.1\%$, and $87.2\%$ of a run's generations elapse after its best construction has appeared, and the best configurations reach their records after just $3.12$, $0.37$, and $5.81$ of the $24$ available hours. Validity is not the bottleneck either: guillotine proposals are usually repairable (median valid yield $86.7\%$, versus $11.1\%$ and $26.0\%$ for the two integrality-gap problems) yet still plateau, so the model is producing legal but not useful novelty.  Taken together, these signatures make the plateau a reachability problem rather than a shortage of computation: the search cannot leave the region it first enters, and the intervention of Section~\ref{sec:seeding} is precisely an attempt to place it in a different region from the outset.

The barriers behind these plateaus\withfull{, ranked by their estimated leverage in \fullver{app:barriers},}{} recur across all three problems: closed representation families whose configurations collapse to a single score, exact witnesses such as LP duals and cut trees that are never fed back as move-level guidance, archive collapse under repeated fresh-model restarts, and the absence of a composition operator that could scale a certified witness while preserving its ratio. 

If the plateau were caused by limited model capacity rather than by the search landscape, a larger transformer of the kind used in the original PatternBoost \cite{charton2024patternboost} should help.  We tested this directly on unit-square stabbing by rerunning representative configurations with a scaled transformer of about $1.82$M parameters in place of the compact $244$K-parameter model, keeping everything else fixed.  We found no clear advantage from the larger model: as Table~\ref{tab:scaled} shows, it matched the compact model on the coordinate-mutation configuration and did slightly worse on the other two.  The scaled model also trains far more slowly and completes only about one third as many search generations under the same wall-clock budget.  Its one potential advantage is per-generation efficiency, since it occasionally reaches a useful construction in fewer generations, but this does not compensate for its slower generations in wall-clock terms.  The evidence therefore points to the geometric search landscape, not transformer capacity, as the binding constraint, and the compact transformer is the better use of a fixed compute budget.

\begin{table}[t]
\caption{Best exact unit-square stabbing gap reached by the compact ($244$K-parameter) and scaled ($1.82$M-parameter) transformers on three representative configurations, under the same wall-clock budget.}
\label{tab:scaled}
\centering
\small
\begin{tabular}{@{}lcc@{}}
\toprule
Configuration & Compact & Scaled \\
\midrule
Coordinate mutation & $1.5000$ & $1.5000$ \\
Symmetry hill climb & $1.5000$ & $1.4286$ \\
Escape search & $1.4400$ & $1.4118$ \\
\bottomrule
\end{tabular}
\end{table}

\section{Human-in-the-loop: an expert-structured staircase construction}
\label{sec:seeding}

\paragraph{Structuring the search space.}\ 
The diagnosis of Section~\ref{sec:aggregate} is that the automated loop is confined to the region it starts in.  Much of the reason is \emph{what it samples}: the neutral search draws square coordinates directly, and among independently placed squares a coupled, provably hard pattern of square--line incidences is astronomically unlikely to arise.  This is where a human enters the loop, not to hand-build the final construction but to change what the search samples.  For unit-square stabbing the vehicle is a \emph{staircase} construction: squares are never sampled independently but generated as products $I\times J$ of intervals from a one-dimensional family, which forces exactly the shared-line incidences that make stabbing hard, and the local moves become scans over the interval starts, the common length, the level count and assignments, the support points, and shifted staircase variants, in place of coordinate jitter on squares.  A line-stabbing lower bound then follows from a proof template on the one-dimensional part, as we now describe.

\paragraph{The staircase formulation.}\ 
\emph{One-dimensional interval layer.}  The construction begins with a finite ordered set of rational support points $P=\{p_1<p_2<\cdots<p_m\}\subset\mathbb{Q}$ and a rational probability measure $\mu:P\to\mathbb{Q}_{\ge 0}$ with $\sum_{p\in P}\mu(p)=1$.  A candidate specifies a common interval length $L$ and a finite family of intervals $\mathcal{I}=\mathcal{I}_0\sqcup\mathcal{I}_1\sqcup\cdots\sqcup\mathcal{I}_{q-1}$ partitioned into ordered \emph{levels}; every interval has the same length $L$ after scaling, and only the starting positions and level labels vary.  For each level $r$ the verifier computes a certified lower bound
\[
a_r\ \le\ \min_{I\in\mathcal{I}_r}\ \mu(I\cap P),
\]
the minimum one-dimensional measure captured by every interval in that level.  The search variables are the interval starts, the level assignments, the common length, and the support weights that make these inequalities strong.

\emph{Product lift to squares.}  The two-dimensional family is formed by products: for $I\in\mathcal{I}_r$ and $J\in\mathcal{I}_s$ the square $Q_{I,J}=I\times J$ is included whenever the level pair is admissible, that is, whenever $a_r+a_s\ge 1$ for the relevant pairs $(r,s)$ below the target threshold.  Because $I$ and $J$ share the length $L$, each $Q_{I,J}$ is an axis-parallel unit square after a uniform rescaling.  A vertical line $x=t$ meets $Q_{I,J}$ exactly when $t\in I$, and a horizontal line $y=t$ exactly when $t\in J$, so stabbing the squares reduces to hitting the interval family on each axis.

\emph{Lower-bound certificate.}  Fix a target integer $T$.  For any split into $u$ vertical and $v$ horizontal lines with $u+v<T$, the staircase levels are arranged so that the vertical lines miss some interval $I\in\mathcal{I}_r$ and the horizontal lines miss some interval $J\in\mathcal{I}_s$ with $(r,s)$ admissible.  The square $Q_{I,J}$ is then present and unstabbed, so no set of fewer than $T$ lines can stab the whole family, which proves $\tau(\mathcal{S})\ge T$.  The fractional optimum $\tau^*(\mathcal{S})$ is computed by the standard LP over the compressed critical-line universe, and the construction is tuned so that $\tau(\mathcal{S})/\tau^*(\mathcal{S})=14/9$.

\paragraph{Results.}\ 
Two observations follow.  The first is negative and instructive: seeding the neutral search with uniformly random instances does not help, because random instances are combinatorially monotonous and leave the search in the same region as before.  A deliberately structured search space behaves very differently.  Working in the staircase formulation, the search reaches a certified unit-square stabbing gap of $\tau/\tau^*=14/9\approx 1.555$.  This exceeds both the best previously known lower bound of $3/2$ and the automated search's best of $20/13$ (Section~\ref{sec:aggregate}).  The gain comes from the structure rather than from any injected answer: the levels, the rational measure, and the product rule supply the proof template, and the search optimizes within it.

\paragraph{The difficulty of producing structure.}\ 
The reason such structure has to come from a human is that the obstructions that force a large gap are hard to manufacture automatically.  For unit-square stabbing, a neutral coordinate search must invent both the geometry and its proof of hardness at once from raw square positions; the staircase formulation removes that double burden.  The same difficulty recurs, often more sharply, elsewhere: for MISR a large gap wants a dense conflict structure, for which a high chromatic number of the intersection graph is a natural proxy, yet rectangle intersection graphs of chromatic number $5$ or $6$ are rare and delicate to build while random ones are almost always low-chromatic, so neither hand construction nor uniform sampling readily supplies them.  Expert structure such as the staircase is valuable precisely because it makes these otherwise-elusive obstructions native to the search space.

\paragraph{Interpretation.}\ 
Modest expert input, supplied precisely where the automated loop stalls, is enough to surpass the best known lower bound for unit-square stabbing, producing a construction that neither the automated search nor the expert reached alone.  %This is the paper's central claim in miniature: machine learning and human expertise are complementary here, the search covering ground that would be tedious by hand and the expert supplying the structural template the search cannot generate for itself.
The same complementarity appears downstream of the search as well as upstream of it: manual post-processing reduces the automated $20/13$ witness from twenty squares to eleven without changing its score, and the $3/2$ MISR construction extends by hand into an infinite family.  In each case, the search supplied a seed that a human could compress or generalize.%Section~\ref{sec:discussion} develops this into a broader view of learning-guided discovery as a partnership.

\section{Conclusion and future directions}
\label{sec:discussion}

Extremal constructions are still found largely by hand.  We showed that a learning-guided search can find them automatically, provided the representation, local search, and surrogate are designed together: it reaches a gap of $3/2$ for MISR, a destroyed fraction of $1/3$ for guillotine cutting, and $20/13\approx1.54$ for unit-square stabbing, the last exceeding the best previously known lower bound; the staircase formulation raises it to $14/9\approx1.555$.  Every construction carries an
independently re-verified certificate.  Two lessons follow: good configurations are isolated in the design space, so no component succeeds alone, and automated search plateaus for a structural rather than a computational reason, since it cannot leave the region it starts in.

Three directions follow from the plateau.  An archive organized by structural features rather than by score would retain candidates that score poorly today but carry the seeds of a better construction; composition operators that combine certified obstructions while preserving their ratio would let a small witness be scaled rather than rediscovered; and the exact structure our solvers already compute -- LP duals for MISR, incidences for stabbing, cut trees for guillotine hardness -- could guide moves directly. The pipeline applies to any extremal question with an exactly computable, certifiable objective \cite{pachwoeginger1990,felsner2013}.  The division
of labor that works is across human and machine: the search covers ground that would be tedious by hand, the expert supplies the template it cannot invent.

\section{Reproducibility and artifacts}
\label{sec:repro}

The code, certificates, and analysis scripts are archived at \url{https://doi.org/10.5281/zenodo.23321041} under CC~BY~4.0.  Every certificate carries a canonical hash, and a verifier independent of the training code recomputes its exact score, so every reported value can be checked directly. The artifact also contains a reduced smoke-test that verifies a small subset of certificates in minutes and a full reproduction script for the main tables. The search used Python~$3.11$, NumPy~$2.0.2$, SciPy~$1.13.1$, and PyTorch~$2.8.0$.

\section*{Acknowledgments}

Dr. Ray's work is supported in part by the NYUAD Center for Interdisciplinary Data Science \& AI (CIDSAI) funded by Tamkeen under the NYUAD Research Institute Award CG016. The research of Dr. Ajwani was supported by Taighde Éireann – Research Ireland, under Grant number [23/RC/13506] at Rinn Artificial Intelligence and Grant number 18/CRT/6183. Aimen Taha's research is supported by a grant from Taighde Éireann – Research Ireland under Grant number 18/CRT/6049. For the purpose of Open Access, the authors have applied a CC BY public copyright licence to any Author Accepted Manuscript version arising from this submission.

The authors used Claude Opus~4.8 and ChatGPT~5.5 to assist with code generation and with the writing of this paper.  All constructions obtained with this assistance were verified using the exact solvers described here, and the authors reviewed and edited every part of the text.%and every part of the text was reviewed and edited by the authors.

\iffullversion
\clearpage
\appendix

\section{Problem illustrations and hard examples}
\label{app:examples}

This appendix illustrates the three problems studied in the paper. Figure~\ref{fig:examples} in the main text shows a small instance of each problem together with an optimal solution, while Figures~\ref{fig:misr-hard}, \ref{fig:sqstb-hard}, and \ref{fig:guil-hard} show small, hand-checkable \emph{hard} instances whose relaxations are far from integral. These hard instances were discovered by domain experts and are not a contribution of this paper. Rather, our goal is to develop methods that can discover such hard instances automatically, and with larger gaps.

\begin{figure*}[!htb]
\centering
\includegraphics[width=0.3\textwidth]{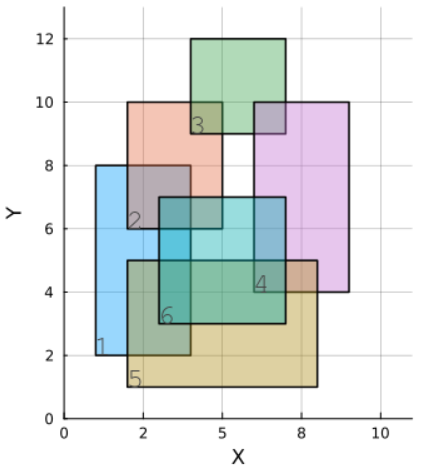}\hfill
\includegraphics[width=0.3\textwidth]{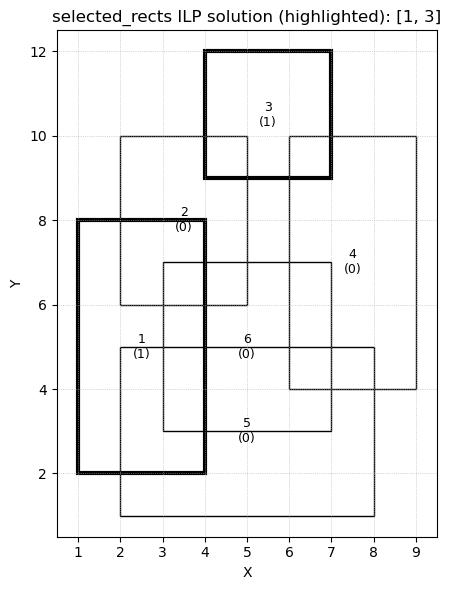}\hfill
\includegraphics[width=0.3\textwidth]{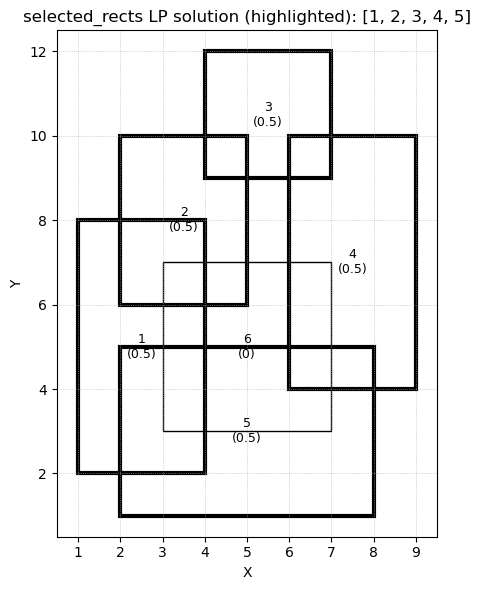}
\caption{A hard MISR instance built around a $5$-cycle (left).  The integer optimum selects two pairwise-disjoint rectangles, $\alpha=2$ (center), whereas the clique LP places weight $1/2$ on all five rectangles of the cycle, $\alpha^*=5/2$ (right), giving an integrality gap of $5/4$.}
\label{fig:misr-hard}
\end{figure*}

\begin{figure*}[!htb]
\centering
\includegraphics[width=0.3\textwidth]{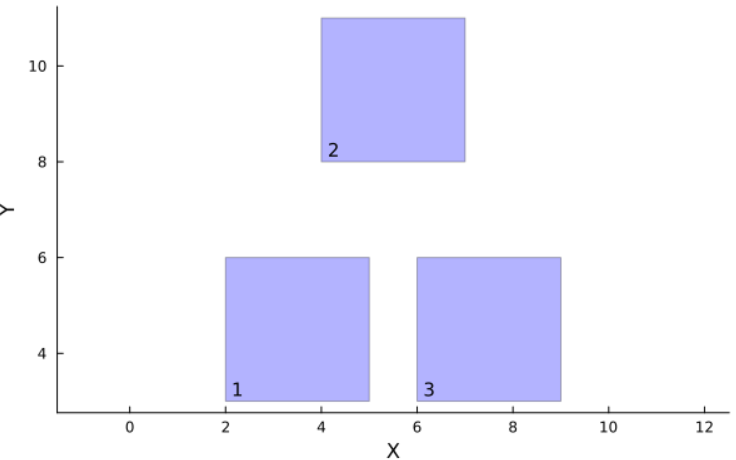}\hfill
\includegraphics[width=0.3\textwidth]{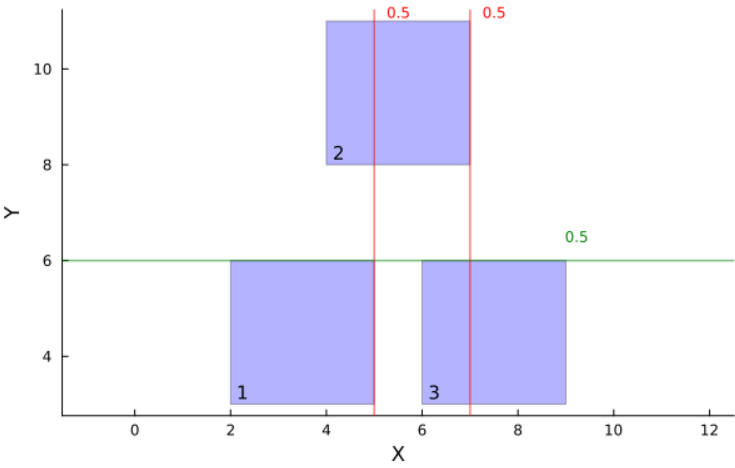}\hfill
\includegraphics[width=0.3\textwidth]{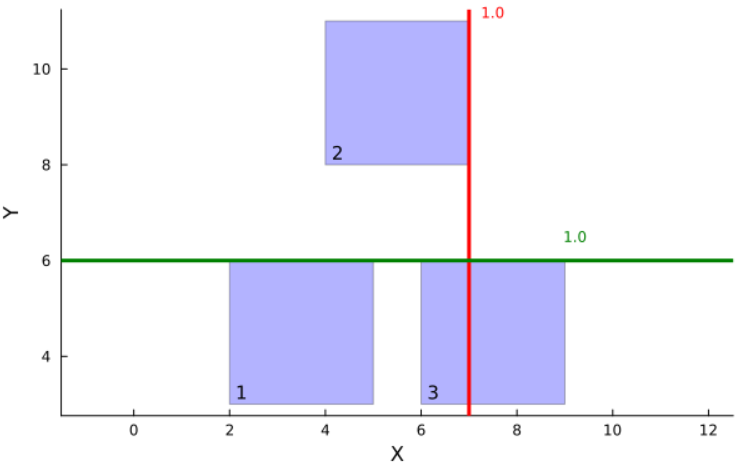}
\caption{A hard unit-square stabbing instance of three squares (left).  The LP places weight $1/2$ on three lines, $\tau^*=3/2$ (center), while any integer stabbing needs two lines, $\tau=2$ (right), giving an integrality gap of $4/3$~\cite{kovalevaspieksma2006}.}
\label{fig:sqstb-hard}
\end{figure*}

\begin{figure*}[!htb]
\centering
\includegraphics[width=0.42\textwidth]{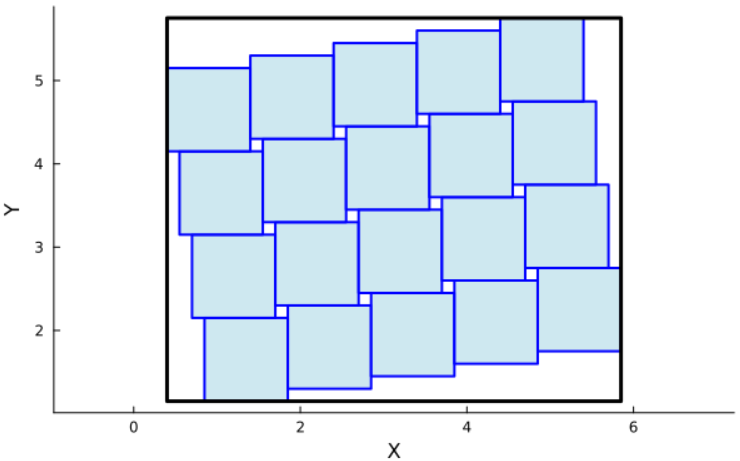}\hfill
\includegraphics[width=0.42\textwidth]{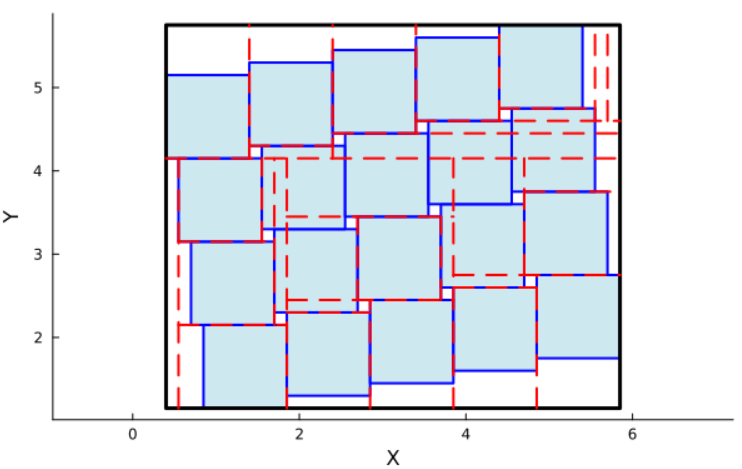}
\caption{A hard guillotine instance: a staggered ``brick-wall'' packing of squares (left).  Because the rows are offset, no full horizontal or vertical cut cleanly separates a row or column, so any recursive guillotine sequence (an optimal one shown by the dashed cuts, right) must destroy a large fraction of the squares~\cite{abed2015}.}
\label{fig:guil-hard}
\end{figure*}

\section{Complete per-configuration results}
\label{app:all81}

Tables~\ref{tab:all81-misr}, \ref{tab:all81-us}, and \ref{tab:all81-guill} report every configuration's best exact score, the gain over its initial construction, the training epoch and elapsed time at which the best appeared, the post-best plateau fraction, and the object count; the top-scoring configuration in each table is shown in bold, and the shortened identifiers are the component names of Table~\ref{tab:components}.  Table~\ref{tab:best} summarizes them.

\begin{table*}[t]
\caption{Complete per-configuration results for MISR ($27$ configurations).  Best is the exact integrality gap $\alpha^*/\alpha$; Gain is the improvement over the initial construction; Epoch and h are the cumulative training epoch and elapsed hours at the best construction; Plat.\ is the post-best plateau fraction; $n$ is the object count.  The top-scoring configuration is highlighted in bold.}
\label{tab:all81-misr}
\centering
\scriptsize
\newcommand{\seqcode}[1]{\begingroup\def\_{\char`\_\discretionary{}{}{}}\texttt{#1}\endgroup}
\renewcommand{\code}[1]{\seqcode{#1}}
\begin{tabular}{@{}p{0.135\textwidth}p{0.115\textwidth}p{0.135\textwidth}rrrrrr@{}}
\toprule
Representation & Local search & Surrogate & Best & Gain & Epoch & h & Plat.\,(\%) & $n$ \\
\midrule
\code{fixed\_sym\_rect} & \code{coeff\_pivot} & \code{exact\_lp\_gap} & 1.2500 & 0.2500 & 564 & 0.42 & 92.9 & 12 \\
\code{fixed\_sym\_rect} & \code{coeff\_pivot} & \code{conflict\_proxy} & 1.2500 & 0.2500 & 2937 & 4.02 & 64.5 & 12 \\
\code{fixed\_sym\_rect} & \code{coeff\_pivot} & \code{tf\_exact\_gap} & 1.2500 & 0.2500 & 4770 & 8.90 & 42.7 & 12 \\
\code{fixed\_sym\_rect} & \code{seq\_pivot} & \code{exact\_lp\_gap} & 1.3333 & 0.3333 & 2256 & 9.12 & 73.0 & 12 \\
\code{fixed\_sym\_rect} & \code{seq\_pivot} & \code{conflict\_proxy} & 1.0000 & 0.0000 & 0 & 0.00 & 100.0 & 12 \\
\code{fixed\_sym\_rect} & \code{seq\_pivot} & \code{tf\_exact\_gap} & 1.2500 & 0.0000 & 0 & 0.00 & 100.0 & 12 \\
\code{fixed\_sym\_rect} & \code{symmetry\_cross} & \code{exact\_lp\_gap} & 1.3333 & 0.3333 & 12 & 0.30 & 99.8 & 12 \\
\code{fixed\_sym\_rect} & \code{symmetry\_cross} & \code{conflict\_proxy} & 1.2500 & 0.2500 & 3 & 0.00 & 99.9 & 12 \\
\code{fixed\_sym\_rect} & \code{symmetry\_cross} & \code{tf\_exact\_gap} & 1.2500 & 0.2500 & 3 & 0.00 & 99.9 & 12 \\
\addlinespace
\code{qp\_rect} & \code{coeff\_pivot} & \code{exact\_lp\_gap} & 1.3333 & 0.3333 & 75 & 0.29 & 98.6 & 8 \\
\code{qp\_rect} & \code{coeff\_pivot} & \code{conflict\_proxy} & 1.3333 & 0.3333 & 192 & 0.33 & 97.2 & 8 \\
\code{qp\_rect} & \code{coeff\_pivot} & \code{tf\_exact\_gap} & 1.3333 & 0.3333 & 1104 & 2.11 & 84.1 & 8 \\
\code{qp\_rect} & \code{seq\_pivot} & \code{exact\_lp\_gap} & 1.3333 & 0.3333 & 510 & 1.19 & 92.2 & 8 \\
\code{qp\_rect} & \code{seq\_pivot} & \code{conflict\_proxy} & 1.3333 & 0.3333 & 1494 & 3.02 & 79.2 & 8 \\
\code{qp\_rect} & \code{seq\_pivot} & \code{tf\_exact\_gap} & 1.3333 & 0.3333 & 336 & 0.60 & 95.2 & 8 \\
\code{qp\_rect} & \code{symmetry\_cross} & \code{exact\_lp\_gap} & 1.2000 & 0.1167 & 0 & 0.02 & 99.9 & 12 \\
\code{qp\_rect} & \code{symmetry\_cross} & \code{conflict\_proxy} & 1.2000 & 0.2000 & 0 & 0.02 & 99.9 & 12 \\
\code{qp\_rect} & \code{symmetry\_cross} & \code{tf\_exact\_gap} & 1.3000 & 0.2167 & 63 & 0.91 & 95.8 & 12 \\
\addlinespace
\code{triangle\_free} & \code{coeff\_pivot} & \code{exact\_lp\_gap} & 1.4000 & 0.4000 & 4782 & 14.48 & 25.1 & 14 \\
\code{triangle\_free} & \code{coeff\_pivot} & \code{conflict\_proxy} & 1.3333 & 0.3333 & 51 & 0.05 & 99.3 & 8 \\
\code{triangle\_free} & \code{coeff\_pivot} & \code{tf\_exact\_gap} & 1.3750 & 0.3750 & 156 & 0.15 & 97.5 & 11 \\
\textbf{\code{triangle\_free}} & \textbf{\code{seq\_pivot}} & \textbf{\code{exact\_lp\_gap}} & \textbf{1.5000} & \textbf{0.3750} & \textbf{1860} & \textbf{3.12} & \textbf{70.7} & \textbf{12} \\
\code{triangle\_free} & \code{seq\_pivot} & \code{conflict\_proxy} & 1.1250 & 0.1250 & 66 & 0.06 & 99.0 & 9 \\
\code{triangle\_free} & \code{seq\_pivot} & \code{tf\_exact\_gap} & 1.3750 & 0.3750 & 231 & 0.22 & 96.4 & 11 \\
\code{triangle\_free} & \code{symmetry\_cross} & \code{exact\_lp\_gap} & 1.2000 & 0.2000 & 0 & 0.00 & 100.0 & 12 \\
\code{triangle\_free} & \code{symmetry\_cross} & \code{conflict\_proxy} & 1.2000 & 0.2000 & 0 & 0.00 & 100.0 & 12 \\
\code{triangle\_free} & \code{symmetry\_cross} & \code{tf\_exact\_gap} & 1.2000 & 0.2000 & 0 & 0.01 & 100.0 & 12 \\
\bottomrule
\end{tabular}
\end{table*}

\begin{table*}[t]
\caption{Complete per-configuration results for unit-square stabbing ($27$ configurations).  Best is the exact integrality gap $\tau/\tau^*$; the remaining columns are as in Table~\ref{tab:all81-misr}.  The top-scoring configuration is highlighted in bold.}
\label{tab:all81-us}
\centering
\scriptsize
\newcommand{\seqcode}[1]{\begingroup\def\_{\char`\_\discretionary{}{}{}}\texttt{#1}\endgroup}
\renewcommand{\code}[1]{\seqcode{#1}}
\begin{tabular}{@{}p{0.135\textwidth}p{0.115\textwidth}p{0.135\textwidth}rrrrrr@{}}
\toprule
Representation & Local search & Surrogate & Best & Gain & Epoch & h & Plat.\,(\%) & $n$ \\
\midrule
\code{fixed\_sym\_grid} & \code{coord\_mut} & \code{exact\_stab\_gap} & 1.5000 & 0.3333 & 1896 & 8.36 & 78.3 & 20 \\
\code{fixed\_sym\_grid} & \code{coord\_mut} & \code{greedy\_lp} & 1.3043 & 0.3043 & 8526 & 23.32 & 1.6 & 20 \\
\code{fixed\_sym\_grid} & \code{coord\_mut} & \code{incidence\_stats} & 1.2766 & 0.2766 & 249 & 0.18 & 97.0 & 20 \\
\code{fixed\_sym\_grid} & \code{sqstab\_hill} & \code{exact\_stab\_gap} & 1.3636 & 0.2098 & 2322 & 5.56 & 63.4 & 20 \\
\code{fixed\_sym\_grid} & \code{sqstab\_hill} & \code{greedy\_lp} & 1.3333 & 0.3333 & 2451 & 4.87 & 64.1 & 20 \\
\code{fixed\_sym\_grid} & \code{sqstab\_hill} & \code{incidence\_stats} & 1.2857 & 0.2857 & 2481 & 4.88 & 63.2 & 20 \\
\code{fixed\_sym\_grid} & \code{symmetry\_cross} & \code{exact\_stab\_gap} & 1.5000 & 0.3462 & 408 & 2.06 & 90.8 & 20 \\
\code{fixed\_sym\_grid} & \code{symmetry\_cross} & \code{greedy\_lp} & 1.5000 & 0.5000 & 309 & 2.66 & 92.9 & 20 \\
\code{fixed\_sym\_grid} & \code{symmetry\_cross} & \code{incidence\_stats} & 1.5000 & 0.3750 & 183 & 0.92 & 95.3 & 20 \\
\addlinespace
\code{line\_sq\_inc} & \code{coord\_mut} & \code{exact\_stab\_gap} & 1.5000 & 0.5000 & 3915 & 11.18 & 40.2 & 21 \\
\code{line\_sq\_inc} & \code{coord\_mut} & \code{greedy\_lp} & 1.2500 & 0.1071 & 4587 & 16.21 & 21.8 & 24 \\
\code{line\_sq\_inc} & \code{coord\_mut} & \code{incidence\_stats} & 1.3333 & 0.3333 & 873 & 1.01 & 87.3 & 9 \\
\code{line\_sq\_inc} & \code{sqstab\_hill} & \code{exact\_stab\_gap} & 1.5000 & 0.5000 & 3003 & 9.29 & 48.9 & 21 \\
\code{line\_sq\_inc} & \code{sqstab\_hill} & \code{greedy\_lp} & 1.3636 & 0.2208 & 2103 & 6.05 & 62.4 & 24 \\
\code{line\_sq\_inc} & \code{sqstab\_hill} & \code{incidence\_stats} & 1.3333 & 0.3333 & 2781 & 7.03 & 54.7 & 15 \\
\code{line\_sq\_inc} & \code{symmetry\_cross} & \code{exact\_stab\_gap} & 1.4644 & 0.3532 & 273 & 11.67 & 58.5 & 28 \\
\code{line\_sq\_inc} & \code{symmetry\_cross} & \code{greedy\_lp} & 1.4141 & 0.4141 & 201 & 1.52 & 56.2 & 20 \\
\code{line\_sq\_inc} & \code{symmetry\_cross} & \code{incidence\_stats} & 1.4268 & 0.4268 & 615 & 20.80 & 8.7 & 20 \\
\addlinespace
\code{sqstab\_grid} & \code{coord\_mut} & \code{exact\_stab\_gap} & 1.5000 & 0.3571 & 117 & 0.35 & 98.5 & 14 \\
\code{sqstab\_grid} & \code{coord\_mut} & \code{greedy\_lp} & 1.3636 & 0.2208 & 6777 & 23.68 & 0.8 & 20 \\
\code{sqstab\_grid} & \code{coord\_mut} & \code{incidence\_stats} & 1.3158 & 0.1158 & 147 & 0.14 & 98.0 & 10 \\
\code{sqstab\_grid} & \code{sqstab\_hill} & \code{exact\_stab\_gap} & 1.5000 & 0.3571 & 375 & 0.46 & 94.8 & 19 \\
\code{sqstab\_grid} & \code{sqstab\_hill} & \code{greedy\_lp} & 1.3462 & 0.3462 & 3087 & 6.68 & 55.5 & 22 \\
\code{sqstab\_grid} & \code{sqstab\_hill} & \code{incidence\_stats} & 1.3333 & 0.3333 & 0 & 0.00 & 100.0 & 9 \\
\textbf{\code{sqstab\_grid}} & \textbf{\code{symmetry\_cross}} & \textbf{\code{exact\_stab\_gap}} & \textbf{1.5385} & \textbf{0.4332} & \textbf{54} & \textbf{0.37} & \textbf{98.2} & \textbf{20} \\
\code{sqstab\_grid} & \code{symmetry\_cross} & \code{greedy\_lp} & 1.5000 & 0.5000 & 39 & 1.41 & 98.7 & 20 \\
\code{sqstab\_grid} & \code{symmetry\_cross} & \code{incidence\_stats} & 1.4407 & 0.4407 & 342 & 2.39 & 87.7 & 20 \\
\bottomrule
\end{tabular}
\end{table*}

\begin{table*}[t]
\caption{Complete per-configuration results for guillotine hardness ($27$ configurations).  Best is the exact destroyed fraction $1-\save/n$; the remaining columns are as in Table~\ref{tab:all81-misr}.  The top-scoring configuration is highlighted in bold.}
\label{tab:all81-guill}
\centering
\scriptsize
\newcommand{\seqcode}[1]{\begingroup\def\_{\char`\_\discretionary{}{}{}}\texttt{#1}\endgroup}
\renewcommand{\code}[1]{\seqcode{#1}}
\begin{tabular}{@{}p{0.155\textwidth}p{0.115\textwidth}p{0.095\textwidth}rrrrrr@{}}
\toprule
Representation & Local search & Surrogate & Best & Gain & Epoch & h & Plat.\,(\%) & $n$ \\
\midrule
\code{fixed\_sym\_pack} & \code{pack\_resize} & \code{depth\_dp} & 0.2500 & 0.0833 & 24 & 0.17 & 99.1 & 12 \\
\code{fixed\_sym\_pack} & \code{pack\_resize} & \code{first\_cut} & 0.2500 & 0.0833 & 42 & 0.42 & 97.1 & 12 \\
\code{fixed\_sym\_pack} & \code{pack\_resize} & \code{k\_subset} & 0.2500 & 0.0833 & 15 & 0.17 & 99.2 & 12 \\
\code{fixed\_sym\_pack} & \code{symmetry\_cross} & \code{depth\_dp} & 0.1667 & 0.0000 & 0 & 13.24 & 43.7 & 12 \\
\code{fixed\_sym\_pack} & \code{symmetry\_cross} & \code{first\_cut} & 0.1667 & 0.0000 & 0 & 8.90 & 62.7 & 12 \\
\code{fixed\_sym\_pack} & \code{symmetry\_cross} & \code{k\_subset} & 0.1667 & 0.0000 & 0 & 24.01 & 0.0 & 12 \\
\code{fixed\_sym\_pack} & \code{witness\_break} & \code{depth\_dp} & 0.2500 & 0.0833 & 6 & 0.06 & 99.7 & 12 \\
\code{fixed\_sym\_pack} & \code{witness\_break} & \code{first\_cut} & 0.2500 & 0.0833 & 3 & 0.02 & 99.9 & 12 \\
\code{fixed\_sym\_pack} & \code{witness\_break} & \code{k\_subset} & 0.2500 & 0.0833 & 12 & 0.09 & 99.5 & 12 \\
\addlinespace
\code{rect\_direct} & \code{pack\_resize} & \code{depth\_dp} & 0.2727 & 0.1061 & 345 & 1.92 & 89.5 & 11 \\
\code{rect\_direct} & \code{pack\_resize} & \code{first\_cut} & 0.2500 & 0.1324 & 2514 & 17.91 & 19.3 & 8 \\
\code{rect\_direct} & \code{pack\_resize} & \code{k\_subset} & 0.2353 & 0.0924 & 306 & 16.72 & 28.3 & 17 \\
\code{rect\_direct} & \code{symmetry\_cross} & \code{depth\_dp} & 0.1667 & 0.0000 & 0 & 24.01 & 0.0 & 12 \\
\code{rect\_direct} & \code{symmetry\_cross} & \code{first\_cut} & 0.1667 & 0.0000 & 0 & 24.01 & 0.0 & 12 \\
\code{rect\_direct} & \code{symmetry\_cross} & \code{k\_subset} & 0.1667 & 0.0000 & 0 & 24.01 & 0.0 & 12 \\
\textbf{\code{rect\_direct}} & \textbf{\code{witness\_break}} & \textbf{\code{depth\_dp}} & \textbf{0.3333} & \textbf{0.2083} & \textbf{858} & \textbf{5.81} & \textbf{74.1} & \textbf{9} \\
\code{rect\_direct} & \code{witness\_break} & \code{first\_cut} & 0.2727 & 0.0727 & 696 & 3.06 & 80.9 & 11 \\
\code{rect\_direct} & \code{witness\_break} & \code{k\_subset} & 0.2500 & 0.1250 & 66 & 2.74 & 87.2 & 8 \\
\addlinespace
\code{obstruction\_grammar} & \code{pack\_resize} & \code{depth\_dp} & 0.2500 & 0.0000 & 0 & 0.00 & 100.0 & 8 \\
\code{obstruction\_grammar} & \code{pack\_resize} & \code{first\_cut} & 0.2500 & 0.0000 & 0 & 0.00 & 100.0 & 8 \\
\code{obstruction\_grammar} & \code{pack\_resize} & \code{k\_subset} & 0.2500 & 0.0000 & 0 & 0.00 & 100.0 & 8 \\
\code{obstruction\_grammar} & \code{symmetry\_cross} & \code{depth\_dp} & 0.1667 & 0.0000 & 0 & 0.00 & 100.0 & 12 \\
\code{obstruction\_grammar} & \code{symmetry\_cross} & \code{first\_cut} & 0.1667 & 0.0000 & 0 & 0.00 & 100.0 & 12 \\
\code{obstruction\_grammar} & \code{symmetry\_cross} & \code{k\_subset} & 0.1667 & 0.0000 & 0 & 0.00 & 100.0 & 12 \\
\code{obstruction\_grammar} & \code{witness\_break} & \code{depth\_dp} & 0.2667 & 0.0167 & 669 & 8.10 & 66.2 & 15 \\
\code{obstruction\_grammar} & \code{witness\_break} & \code{first\_cut} & 0.2727 & 0.0227 & 468 & 2.54 & 85.8 & 11 \\
\code{obstruction\_grammar} & \code{witness\_break} & \code{k\_subset} & 0.2609 & 0.0109 & 510 & 22.87 & 4.6 & 23 \\
\bottomrule
\end{tabular}
\end{table*}

\section{Component-slice figures}
\label{app:slices}

Figures~\ref{fig:slice-misr} and \ref{fig:slice-guillotine} give the component-slice convergence plots for MISR and guillotine hardness, complementing the unit-square panel shown as Figure~\ref{fig:slice-unitsquare} in Section~\ref{sec:aggregate}.  Each panel fixes two of the three components at their best-performing choice and varies the third, plotting the best exact score found so far against cumulative transformer-training epochs.

\begin{figure*}[t]
\centering
\includegraphics[width=\textwidth]{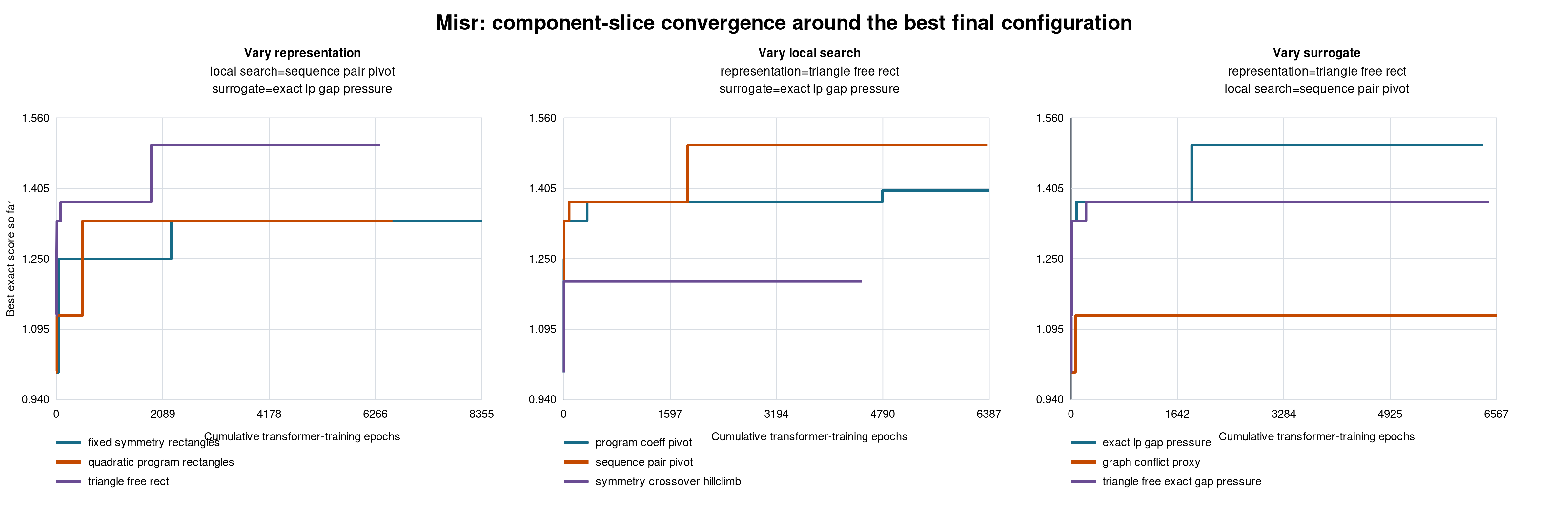}
\caption{Component-slice convergence for MISR, plotting the best exact integrality gap $\alpha^*/\alpha$ found so far.  From left to right the panels vary the representation, the local-search operation, and the surrogate.}
\label{fig:slice-misr}
\end{figure*}

\begin{figure*}[t]
\centering
\includegraphics[width=\textwidth]{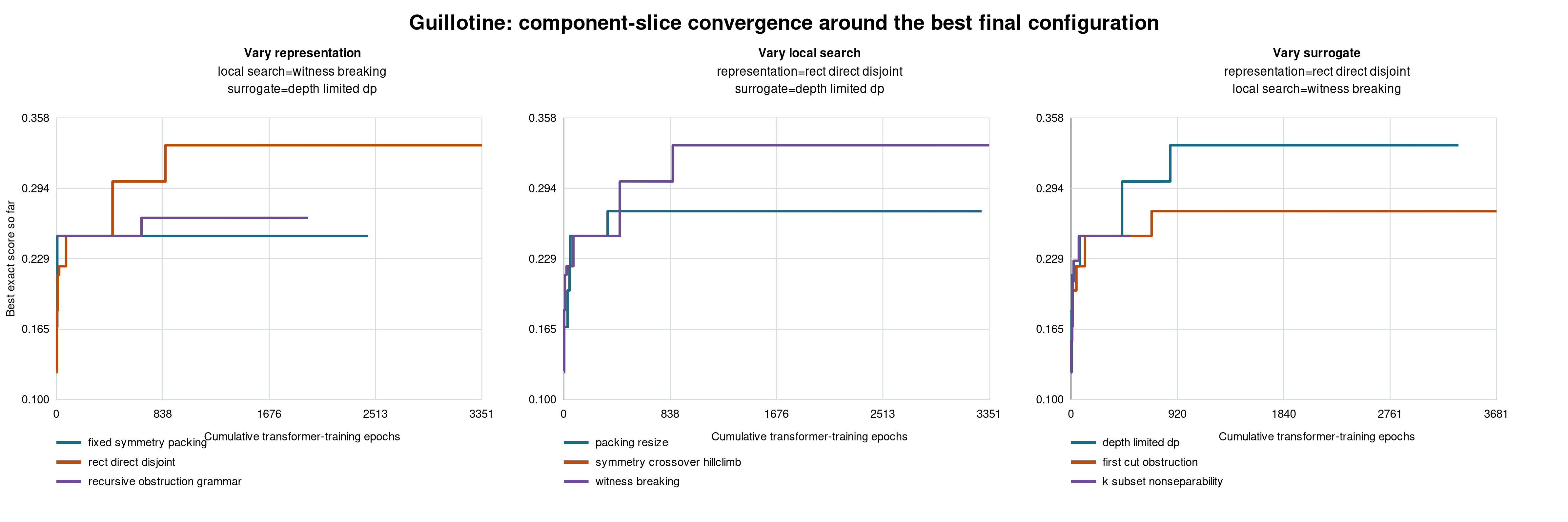}
\caption{Component-slice convergence for guillotine hardness, plotting the best exact destroyed fraction $1-\save/n$ found so far.  From left to right the panels vary the representation, the local-search operation, and the surrogate.}
\label{fig:slice-guillotine}
\end{figure*}

\section{Detailed component descriptions}
\label{app:components-detail}

\renewcommand{\code}[1]{\begingroup\def\_{\char`\_\discretionary{}{}{}}\texttt{#1}\endgroup}

This appendix expands the $27$ component choices of Table~\ref{tab:components} into a fuller description of what each component stores, how it initializes and repairs candidates, and how it moves or ranks them.  Throughout, ``repair'' denotes the map that turns a proposed object back into valid geometry, and only the exact verifier of Section~\ref{sec:problems} determines a reported score.

\subsection{MISR components}\ 

\emph{Representations.}
\code{triangle\_free\_rect} stores a list of integer rectangle coordinates under the constraint that the intersection graph stay triangle-free, that is, no three rectangles share a common point.  Rectangles are added one at a time from a pool that mixes rectangles near existing ones, thin horizontal and vertical bars, and unrestricted random rectangles; a candidate is admissible only if its neighbors in the current intersection graph are pairwise non-intersecting, which is exactly what prevents a triangle.  Repair deletes a member of any triangle that appears and refills to the minimum size.  Because sparse conflict graphs can support large fractional-versus-integer gaps this is a strong but natural restriction, but it cannot reach a construction whose intersection graph must contain a triangle.
\code{quadratic\_program\_rectangles} does not store rectangle coordinates directly; it stores four short bounded integer coefficient vectors, expands each rectangle index into a small quadratic feature vector of its binary digits, and decodes four dot products with those coefficients into a rectangle (through one of three decoders: sorted endpoints, center-and-size, or alternating thin segments).  Changing a single coefficient therefore moves many rectangles at once, giving cheap, globally correlated moves.  Decoded rectangles are cleaned and repaired to a triangle-free family, so many different coefficient programs collapse to the same geometry; the representation favors concise algebraic regularity but can be dominated by its repair step.
\code{fixed\_symmetry\_rectangles} stores direct integer rectangles at a fixed object count, with no triangle-free constraint.  Each family is translated to the origin and reduced to the lexicographically smallest of its eight images under the planar reflections and quarter-turns, so mirror and rotated copies collapse to a single training string.  Repair clamps coordinates, restores the target cardinality, and re-canonicalizes; because the cardinality is fixed, an apparent add or delete may be undone or completed by repair.

\emph{Local-search operations.}
\code{sequence\_pair\_pivot} keeps a historical name (no final representation stores endpoint sequences).  On the triangle-free representation it samples add, replace, nudge, resize, and delete moves with add and replace upweighted, drawing new rectangles from a large triangle-free pool; on the quadratic-program representation it uses the generic coefficient, decoder, and composition neighborhood; and on the fixed-symmetry representation it becomes direct coordinate jitter, resize, duplicate, and delete followed by fixed-cardinality repair.
\code{program\_coeff\_pivot} is most meaningful with the quadratic-program representation, where it changes one coefficient by a small step, refreshes a short coefficient block, switches among the three decoders, or changes the active object count.  Because a single coefficient enters the quadratic feature of many rectangle indices at once, such a move is global and correlated: it can reshape several rectangles together, unlike nudging one endpoint of one rectangle.  With the other two representations there is no coefficient payload to perturb, so it falls back to the safe direct-coordinate neighborhood and is a genuinely distinct coefficient search only in the program panel.
\code{symmetry\_crossover\_hillclimb} is a shared operator that makes $25$ exactly scored proposals per call, drawn as about $55\%$ one-step shifts, $25\%$ random relocations, $10\%$ pairwise exchanges (swapping the vertical intervals of two rectangles, a within-candidate recombination rather than crossover between parents), and $10\%$ planar symmetries.  Every proposal is repaired and exactly scored and is accepted only if the exact objective does not worsen, so the operator is expensive and strongly exploitative and cannot cross a valley by accepting a temporary decrease; paired with the fixed-symmetry representation, the global symmetry proposal canonicalizes back to the same state and is neutral.

\emph{Surrogates.}
\code{exact\_lp\_gap\_pressure} calls the full MISR exact scorer for every candidate it ranks and orders the population primarily by the exact clique-LP-to-integer gap $\alpha^*/\alpha$, the very quantity the study reports, breaking ties among candidates of equal gap in favor of a larger triangle-free lower bound.  It is called a surrogate only because it decides which candidates are retained and trained on, not because it is approximate: it is the most tightly aligned of the three MISR signals, and also the most expensive, since each ranking call solves the same optimization as the verifier and so lowers the number of generations completed within the time budget.
\code{triangle\_free\_exact\_gap\_pressure} first applies a hard filter, pushing any candidate whose intersection graph contains a triangle to the bottom so that it is effectively discarded.  Among the surviving triangle-free candidates it ranks by the exact gap, breaking ties in favor of a stronger triangle-free lower bound and a denser conflict graph (more intersection edges relative to the object count).  Because the rejection happens in the ranking step rather than in the representation, this surrogate can enforce triangle-freeness even when paired with a representation, such as \code{fixed\_symmetry\_rectangles}, that does not.
\code{graph\_conflict\_proxy} avoids solving the clique LP and instead estimates promise from cheaper graph statistics: it rewards a high intersection-edge density relative to the exact independence number, that is, many conflicts packed against a small independent set, adds a term for the maximum clique size, and gives a bonus for triangle-freeness.  The intuition is that large gaps tend to live in graphs that are dense in conflicts yet have a small independent set.  It is cheaper than solving the LP but not a purely linear-time statistic, since computing the independence and clique numbers still needs combinatorial graph routines, and edge density alone need not predict the true clique-LP gap.

\subsection{Unit-square stabbing components}\ 

\emph{Representations.}
\code{line\_square\_incidence} builds a balanced abstract row-and-column layout first and maps it to square anchors, assigning each square an abstract row and column and jittering each axis with modest probability.  The solver-facing state is the list of anchors, with the row and column labels recomputed from the geometry after each move, so the incidence payload guides initialization and structured moves rather than being a separate exact model.  It starts near balanced line-incidence patterns, which makes hard covers more likely but can overproduce easy-to-stab grids.
\code{sqstab\_exact\_grid} stores distinct integer lower-left anchors and one common integer side length, with initial anchors sampled without replacement from a grid and the initial side drawn from $1$ to $4$.  Canonicalization removes translation and duplicates and keeps a common side, so the exact critical-line universe is easy to regenerate and audit.  It is the least indirect of the three square representations, searching the actual coordinates and scale, at the cost of a large, rugged space with many plateaus.
\code{fixed\_symmetry\_grid} stores distinct grid anchors at a fixed object count and a fixed common side (two in the final runs), reduced over all translations and the eight symmetries to a canonical coordinate list.  Cardinality and side length are restored after every proposal, which makes representation comparisons clean at a fixed scale but prevents the search from moving between side-length regimes.

\emph{Local-search operations.}
\code{coord\_mutation} moves one square by at most one grid step, duplicates a nearby square, or deletes one.  Representation-specific additions matter: with the incidence layout it can copy another square's $x$- or $y$-coordinate; with the exact grid it also has line tightening, canonical shifting, a random restart, and a guided neighbor selector; and with the fixed-symmetry grid, fixed-count repair restores cardinality and re-canonicalizes.  It is a plain coordinate mutation only in its simplest pairing.
\code{sqstab\_local\_hillclimb} augments the base moves with stabbing-aware proposals: copying or tightening a square onto another's coordinate (optionally offset by the side length), adding or replacing a unique anchor, dropping a square, shifting to the origin, or restarting with fresh grid points and a resampled side.  After the first proposal it builds a few nearby candidates and, for families up to about $48$ squares, keeps the one with the best exact stabbing key; above that size it switches to the cheap greedy-cover-versus-frequency proxy to control cost.
\code{symmetry\_crossover\_hillclimb} uses the same $25$-proposal shift, relocate, exchange, and symmetry portfolio, exactly scored.  Acceptance is gated first by a nontriviality requirement -- a candidate needing at least four lines is preferred to one with a smaller integer cover -- and then by the exact gap; within the gate only nonworsening proposals are kept.  The gate is an explicit search preference, not a target ratio or injected construction.

\emph{Surrogates.}
\code{greedy\_partial\_lp\_bitset} estimates the gap without solving any LP.  It divides the size of a greedy line cover (an integer numerator) by a crude fractional lower bound, namely the number of squares divided by the largest number of squares any single line stabs; because a line that stabs many squares forces the fractional optimum up, this ratio bounds how good any cover can be.  Incidence sets are stored as bitmasks, so the whole computation is fast, but the estimate can overrate a candidate when the greedy cover is poor for incidental reasons or when the maximum-frequency bound is much weaker than the true fractional optimum.
\code{exact\_stab\_gap\_pressure} calls the exact set-cover and fractional-LP scorer for each candidate and ranks primarily by the exact stabbing gap $\tau/\tau^*$.  Its small tie-breaks prefer, among candidates of equal gap, those with a larger integer cover (more genuinely required lines), a smaller fractional optimum, and a more spread-out LP support, that is, more nonzero line variables, which tends to signal a richer fractional structure.  As with the exact MISR surrogate, ``surrogate'' names its role in population ranking rather than any loss of fidelity, and its cost is that each call solves the same problem as the verifier.
\code{incidence\_statistics} scores from line-incidence summaries alone and never solves a cover.  It takes the greedy cover size, divides it by one plus the spread (standard deviation) of how many squares each critical line stabs, so that balanced, uniform incidence is rewarded, and adds a small bonus for a large number of distinct critical-line patterns.  It is the cheapest square surrogate but is easily misled, since combinatorial incidence diversity need not translate into a large exact gap.

\subsection{Guillotine components}\ 

\emph{Representations.}
\code{rect\_direct\_disjoint} initializes by turning random grid cells into disjoint unit rectangles, a deliberately neutral start that usually yields an easily separable packing.  Repair clamps intervals, removes interior overlaps, refills to the minimum size, and canonicalizes coordinates; the local search is responsible for growing the simple cells into long bars and interlocking packings.
\code{recursive\_obstruction\_grammar} instantiates one or more seven-rectangle spiral obstruction cores, adds disjoint filler rectangles, and canonicalizes.  No certificate or target score is inserted and the coordinates vary (consecutive gaps sampled from small values, with random reflections and axis exchange), but the seven-rectangle spiral topology is a hand-designed prior, so this is an obstruction-family search rather than a neutral coordinate search.
\code{fixed\_symmetry\_packing} samples disjoint rectangles in a dynamic square domain, with roughly $40\%$ long horizontal bars, $40\%$ long vertical bars, and $20\%$ compact rectangles, rejecting overlaps.  The object count is fixed and the family is canonicalized over all symmetries; unlike the grammar it embeds no fixed non-sliceable topology, its only prior being the bar-heavy shape distribution.

\emph{Local-search operations.}
\code{packing\_resize} is a geometry-first neighborhood of move, resize, add, and delete (with add upweighted) followed by disjointness repair.  After the primary move it runs three guided trials that may close a projection gap, add a bridge or bar, replace a rectangle, or nudge an endpoint, keeping a nonworsening candidate under a nonseparability key; this second stage is what makes it more than random packing noise.
\code{witness\_breaking} first finds an easy witness, a guillotine-separable subset of majority size (enumerated for small $n$, sampled otherwise), and then tries to destroy that separation by extending a rectangle across a projection gap or inserting a thin bridging rectangle, alongside bar replacement, cut blockers, and motif copies, again followed by guided nonseparability trials.  It is an adversarial neighborhood that acts on a concrete weakness of the current packing.
\code{symmetry\_crossover\_hillclimb} applies the shared $25$-proposal portfolio -- one-step shifts, relocations that preserve a rectangle's width and height, exchanges of two rectangles' vertical intervals, and global symmetries -- to disjoint packings, repairing each proposal to interior-disjointness and scoring it exactly.  Acceptance is gated first by a nontriviality requirement of at least two destroyed rectangles and then by the exact number destroyed, and no exact-downhill step is ever accepted.  In practice this shared exact-nonworsening rule does not transfer to guillotine hardness: starting from an easily separable packing, none of its nonworsening local moves reaches a harder one, so every guillotine configuration using it stalls at the trivial $1/6$ value (Section~\ref{sec:aggregate}).

\emph{Surrogates.}
\code{first\_cut\_obstruction} scores by how many rectangles the cheapest progressing first cut must cross, that is, how hard it is to make even the first guillotine slice, with small bonuses for a first cut that leaves two comparably sized sides and for a packing whose horizontal and vertical spans are similar.  It is fast and intuitive but poorly aligned with the recursive objective, since a construction can force an expensive first cut and yet split into two easy subproblems, so a high first-cut cost need not mean a high final destruction fraction.
\code{depth\_limited\_dp} runs the recursive-cut dynamic program to depth four, optimistically treating every rectangle still present at the depth limit as saved, and ranks by the implied destroyed fraction plus small balance and span bonuses.  It sees past the first cut while staying cheaper than the full dynamic program, but misses obstructions that surface deeper than four levels.
\code{k\_subset\_nonseparability} is the richest guillotine signal: it measures the fraction of majority-size subsets that are not guillotine-separable (exhaustive for small $n$, sampled otherwise) and combines it with the depth-four estimate, a width-six beam-search estimate, and balance and span terms.  Its weakness is that sampled majority-subset resistance is not the same as the exact destruction fraction, especially at larger $n$.

\section{Prioritized limitations}
\label{app:barriers}

Table~\ref{tab:limits} organizes the barriers behind the plateaus of Section~\ref{sec:aggregate} and ranks them by \emph{leverage}, our estimate of how much removing a barrier could change the reachable constructions rather than how easy it would be to remove.

\begin{table*}[t]
\caption{Prioritized limitations supported by the evidence across all runs.  ``Leverage'' estimates how strongly fixing the limitation could change the reachable constructions, not how easy the fix is.}
\label{tab:limits}
\centering
\small
\begin{tabularx}{\textwidth}{@{}p{0.20\textwidth}>{\raggedright\arraybackslash}X p{0.12\textwidth}@{}}
\toprule
Limitation & Evidence and consequence & Leverage \\
\midrule
Closed representation families & All $9$ symmetry-crossover guillotine configurations finish at $1/6$, and $9$ MISR configurations across several component paths finish at $4/3$.  Search cannot reach a structure its representation and repair cannot express. & Very high \\
Weak exact-witness feedback & Exact primal, dual, stabbing, and cut-tree witnesses are not turned into move-level feedback; learned-model samples yield only $0$, $1$, $1$ strict records. & Very high \\
Archive collapse and model restart & Median duplicate-slot fractions reach $42.6\%$, $36.1\%$, $22.6\%$; each event fits a fresh three-epoch model, so features do not accumulate. & Very high \\
No composition mechanism & Best certificates have only $12$, $20$, $9$ objects; no operator combines two certified blocks while preserving the ratio. & Very high \\
Discrete exact objectives & Small changes usually leave the ratio unchanged; the next useful change needs several coordinated objects, explaining $64$--$98\%$ plateaus. & High \\
Proposal--repair mismatch & Repair guarantees validity but can undo a coordinated proposal; high valid-sample rates coexist with little exact progress. & High \\
Syntactic serialization & The model sees serialized coordinates, not canonical graphs, incidences, duals, or cut trees, so capacity is spent relearning equivalent encodings. & Med.--high \\
\bottomrule
\end{tabularx}
\end{table*}

\section{PatternBoost pseudocode}
\label{app:pseudocode}

The loop of Section~\ref{sec:method}, shown in Figure~\ref{fig:pbloop}, is summarized below.  It is a population-based search: each generation scores the population with the surrogate, exactly audits the elites at a fixed interval, periodically retrains a fresh model on the archive, and refills the population by mutating elites with the configuration's local-search operator and adding a few random newcomers.

\noindent\begin{minipage}{\columnwidth}
\begin{verbatim}
PatternBoost(problem, repr, local, surr):
  pop <- fresh random instances
  archive <- {};  best <- none
  for g in 1..iterations:

    # score population with the surrogate
    scored <- []
    for x in pop:
      x <- repair(repr, x)
      if x seen before: continue
      s <- surrogate(surr, x)
      scored += (s, x);  archive += (s, x)
    elites <- top-k of scored by surrogate

\end{verbatim}
\end{minipage}
\medskip

\noindent\begin{minipage}{\columnwidth}
\begin{verbatim}

    # periodic exact verification
    if g mod exact_every == 0:
      for x in elites:
        c <- exact_score(problem, x)
        if c ok and c.score > best.score:
          save certificate;  best <- c

    next <- elites

    # global learning: train a fresh model
    if g mod train_every == 0 and archive:
      model <- train_model(top of archive)
      for y in sample(model):
        y <- repair(repr, y)
        if y ok and non-trivial:
          next += y

    # local search, then random newcomers
    while |next| < pop_size - num_new:
      p <- next elite (else random)
      next += mutate(local, p)
    next += num_new random instances

    pop <- next[:pop_size]
    checkpoint periodically
  return best
\end{verbatim}
\end{minipage}

\section{Certificate schemas and verification}
\label{app:certs}

Every reported construction is stored as a standalone certificate that records the construction together with the components of its exact score.  A verifier independent of the training code recomputes the exact score from each certificate, and every certificate carries a canonical hash.  The minimal fields per problem are as follows.

\noindent\begin{minipage}{\columnwidth}
\begin{verbatim}
MISR (misr_certificate_v1):
  rectangles       [[x1,x2,y1,y2], ...]
  alpha_int        integer optimum   (alpha)
  alpha_lp         clique-LP optimum (alpha*)
  score            alpha_lp / alpha_int
  maximal_cliques  [[0,3,7], ...]
  integer_solution [0,4,9]
  solver_status    "optimal"
  certificate_hash "..."
\end{verbatim}
\end{minipage}
\medskip

\noindent\begin{minipage}{\columnwidth}
\begin{verbatim}
Unit-square (unit_square_stab_certificate_v1):
  squares          [[x,y], ...]
  critical_lines   {vertical:[...],
                    horizontal:[...]}
  tau_int          integer optimum  (tau)
  tau_lp           set-cover LP opt  (tau*)
  score            tau_int / tau_lp
  integer_lines    ["v3","h5", ...]
  solver_status    "optimal"
  certificate_hash "..."
\end{verbatim}
\end{minipage}

% Column break: the guillotine schema starts the right-hand column, so the
% unit-square and guillotine schemas sit side by side under Table~\ref{tab:limits}.
\newpage

\noindent\begin{minipage}{\columnwidth}
\begin{verbatim}
Guillotine (guillotine_certificate_v1):
  rectangles       [[x1,x2,y1,y2], ...]
  n                number of rectangles
  saved            max preserved by optimal
                   recursive cutting
  destroyed        n - saved
  score            destroyed / n
  dp_states        number of DP states
  optimal_strategy recursive cut tree {...}
  solver_status    "optimal"
  certificate_hash "..."
\end{verbatim}
\end{minipage}

\fi

\end{document}